\documentclass[aps,prb,showpacs,twocolumn,floats,epsfig,pdflatex]{revtex4}
\usepackage{amssymb}
\usepackage{amsbsy}
\usepackage{amsmath}
\usepackage{epsfig}
\usepackage{graphicx}
\usepackage{epsfig,amssymb,amsmath}
\usepackage{color}
\begin{document}
\title{Effect of Magnetic Field and Rashba Spin-Orbit Interaction on the Josephson Tunneling between Superconducting Nanowires}
\author{E.  Nakhmedov$^{1,2}$}
\author{O. Alekperov$^{1}$}
\author{F. Tatardar$^{1,3}$}
\author{Yu. M. Shukrinov$^{4,5}$}
\author{I. Rahmonov$^{4,6}$}
\author{K. Sengupta$^{7}$}

\address{
$^{1}$Institute of Physics of National Academy of Sciences of Azerbaijan, H. Javid ave. 133, Baku, AZ-1143, Azerbaijan\\
$^{2}$ Faculty of Physics, Moscow State University, Baku branch, str. Universitetskaya 1, AZ-1144 Baku, Azerbaijan,\\
$^{3}$Khazar University, Mahsati str. 41, AZ 1096, Baku, Azerbaijan,\\
$^{4}$ BLTP, JINR, Dubna, Moscow region, 141980, Russian Federation,\\
$^{5}$ Dubna State University, Dubna, Moscow region, 141980, Russian Federation,\\
$^{6}$ Umarov Physical Technical Institute, TAS, Dushanbe, 734063 Tajikistan\\
$^{7}$Theoretical Physics Department, Indian Association for the Cultivation of Science, Jadavpur, Kolkata 700 032, India}

\date{\today}

\begin{abstract}

We calculate the Josephson current between two one-dimensional (1D)
nanowires oriented along $x$ with proximity induced $s$-wave
superconducting pairing and separated by a narrow dielectric barrier
in the presence of both Rashba spin-orbit interaction (SOI)
characterized by strength $\alpha$ and Zeeman fields ($h$ along
$\hat z$ and ${\bf B}$ in the $x-y$ plane). We formulate a general
method for computing the Andreev bound states energy which allows us
to obtain analytical expressions for the energy of these states in
several asymptotic cases. We find that in the absence of the
magnetic fields the energy  gap between the Andreev bound states
decreases with increasing Rashba SOI constant leading eventually to
touching of the levels. In the absence of Rashba SOI, the Andreev
bound states depend on the magnetic fields and display oscillatory
behavior with orientational angle of B leading to magneto-Josephson
effect. We also present analytic expressions for the dc Josephson
current charting out their dependence on ${\bf B}$, $h$, and
$\alpha$. We demonstrate the existence of finite spin-Josephson
current in these junctions in the presence of external magnetic
fields and provide analytic expressions for its dependence on
$\alpha$, $\bf B$ and $h$. Finally, we study the AC Josephson effect
in the presence of the SOI (for $|{\bf B}|=h=0$) and an external
radiation and show that the width of the resulting Shapiro steps in
such a system can be tuned by varying $\alpha$. We discuss
experiments which can test our theoretical results.
\end{abstract}
\maketitle
\section{Introduction}
\label{intro}

A recent idea on possible application of topological superconductors
to quantum information processing has attracted both theoretical and
experimental interest \cite{kitaev03, nssf08}. According to this
idea, a quantum information unit, qubit,  can be formed and
propagated by means of Majorana mode (see, for e.g., Ref.\
\onlinecite{alicea12}), localized at the end of a one-dimensional
(1D) chain hosting a topological superconductor \cite{mzfp12,sen1}.
Recent investigations suggest several detection mechanisms of such a
Majorana mode \cite{mzfp12,sen1, ksy04,drmo12, rlf12, dyhl12,
cfgd13, fhmj13} such as an existence of a central peak in the
tunneling current through a topological superconductor (S)- normal
metal (N)junction and fractional period of the Josephson current in
S-N-S junctions. Most recently experimental systems involving 1D
semiconductor wire has been shown to host such modes; the mechanism
of the appearance of such modes arise from the combination of strong
SOI, proximity-induced superconducting gap, chemical potential and
applied Zeeman field in these wires \cite{lsd10, oro10}. Two
Majorana modes in the Josephson junction, formed between two
topological insulator edges or one-dimensional superconducting
nanowires separated by barrier, hybridize resulting in splitting of
the zero energy modes. This splitting energy depends not only on the
phase difference of the two superconductors but also on the relative
direction of the spin polarization at the two side of the junction.

The oscillations of a Josephson current between two such
superconductors separated by insulator or metal as a function
of their phase difference, with $4 \pi$ periodicity instead of a
conventional $2 \pi$ periodicity due to hybridization of Majorana
states was predicted by Kitaev \cite{kitaev01} for a idealized model
of an 1D spinless p-wave superconductor. Following this, Kwon {\it
et al}. \cite{ksy04} proposed that the similar effect can be
observed between quasi-1D or 2D unconventional superconducting
tunnel barrier junctions where the superconductors are separated by
an insulating region, usually modeled by a delta function potential
barrier. These systems did not have SOI or Zeeman field;
Majorana-like modes appeared in such systems from the unconventional
nature of the pairing potential. Further it was realized in Ref.\
\onlinecite{ksy04} that a signature of the fractional Josephson
effect constitutes in having a halved Josephson frequency,
$\omega_J=eV/\hbar$, in the presence of a DC voltage $V$ applied
across the junction. These effects have been interpreted in terms of
the Josephson current being carried by electrons rather than Cooper
pairs \cite{ksy04}. Further, it was shown that a fractional
Josephson effect may be realized at topological insulator edge
\cite{fk08, fk09}. This prediction has later been extended to
different systems \cite{bhm11, jpar11, spa12, pn12, ojanen13,
scpa13, lmay14, clgt15}.

Recent activities have established that a topological insulator with
proximity-induced coupling to a s-wave superconductor exhibits a
superconductivity-magnetism duality \cite{nab08, jpar11, jpar13,
kss12, bphs13, pjpa13}, revealing the fractional periodicity not
only with superconducting phase difference but also with the
orientation of Zeeman magnetic field. In this case, the magnetic
field on one side of the junction rotates in the plane normal to the
direction of an effective magnetic field of the SOI; consequently,
the Majorana-mediated Josephson current reverses sign after $2\pi$
rotation of the magnetic field orientation and reveals an
unconventional $4 \pi$ periodic magneto-Josephson oscillation in
response to variation of the magnetic field orientation in a
topological insulator edge \cite{jpar11, kss12}. Furthermore, a
dissipationless fractional Josephson effect mediated by with $8 \pi$
periodicity has been also predicted \cite{zk14} at the edge of a
quantum spin Hall insulator. The Josephson effect in consisting of
topological superconducting (S) and normal (N) regions, has been
reported in \cite{jpar11,kss12,jpar13, pjpa13}. These works also
reveal a signature of Majorana bound states located at S-N edges,
producing a fractional Josephson current with $4 \pi$ periodicity
\cite{ksy04}.

These previous works in the field have pointed out the importance of
the fractional Josephson and the magneto-Josephson effect at the
edge of junction of topological insulators.  However, the role of
spin-orbit coupling and the external magnetic field have not been
investigated for $1$D superconducting junctions in these earlier
works. In particular, a theoretical formalism for computation of
Andreev bound states which requires an extension of the work of
Ref.\ \onlinecite{ksy04} to systems with SOC and Zeeman fields is
lacking. The development of such a formalism and a systematic study
of its results is the main aim of the present work. To this end, we
study the Josephson effect between two 1D nanowires oriented along
$x$ with proximity induced $s$-wave superconducting pairing and
separated by a narrow dielectric with a Rashba spin-orbit
interaction (SOI) of strength $\alpha$ and Zeeman fields ($h$ along
$\hat z$ and ${\bf B}$ in the $x-y$ plane). A schematic
representation of the proposed setup is shown in Fig.\
\ref{Josephson}.

The main results of our study are as follows. First, we develop a
general method for computing the Andreev bound states energy in
these junctions. Such a method constitutes a generalization of the
method of Ref.\ \onlinecite{ksy04} to junctions with Zeeman magnetic
fields and spin-orbit coupling. Second, using this method, we obtain
analytical expressions for the energy of the Andreev bound states in
several asymptotic cases and discuss their implication on the
Josephson current. For example, we find that in the absence of the
magnetic fields the energy gap between these bound states decreases
with increasing Rashba SOI constant leading eventually to level
touching while in the absence of Rashba SOI,
they  display oscillatory behavior with orientational angle of ${\bf
B}$. Third, we present analytic expressions for the dc Josephson
current charting out their dependence on both ${\bf B}$ and $h$ and
the SOI interaction strength. Fourth, we demonstrate the existence
of finite spin-Josephson current in these junctions in the presence
of external magnetic fields and provide analytic expressions for its
dependence on $\alpha$, $\bf B$ and $h$. Finally, we study the AC
Josephson effect in the presence of the SOI (for $|{\bf B}|=h=0$)
and an external radiation and show that the width of the resulting
Shapiro steps in such a system can be tuned by varying $\alpha$. We
discuss experiments which can test our theoretical results.

The plan of the rest of the paper is as follows. In Sec. \ref{sec2},
we describe the model and present explicit form of Hamiltonian.
The hybridization energy of edge states is calculated in
Sec.\ \ref{sec3}, where several asymptotic expressions for the
Josephson coupling energy are obtained. This is followed by a
discussion of the DC Josephson effect in Sec.\ \ref{sec4}. The AC
Josephson effect in these system and the dependence of the Shapiro
step on SOI strength is studied in Sec.\ \ref{sec5}. Finally we
conclude in Sec.\ \ref{sec6}. Some details of our calculations are
specified in the Appendices.

\section{Model and formulation of the problem}
\label{sec2}

\begin{figure}
\centering
\includegraphics[height=65mm]{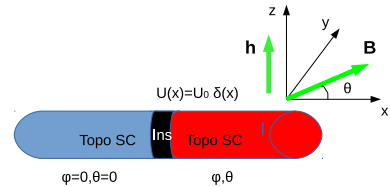}
\caption {Two s-wave superconductors separated with $\delta$-like
dielectric potential under magnetic fields ${\bf h}$ and ${\bf B}$
co-planar and perpendicular to spin-orbit interaction respectively.
The bulk s-wave superconductors which induces superconductivity in
the wires are not shown for clarity.} \label{Josephson}
\end{figure}

We consider a junction of two 1D nanowires with proximity induced
$s$-wave pairing symmetry in the presence of Rashba spin-orbital
interaction and external magnetic fields. The schematic
representation of such a junction is shown in Fig.\ \ref{Josephson}
where the proximate superconductors are not shown for clarity.

In what follows we assume the pairing is induced by two proximate
$s$-wave superconductors which leads to effective pairing potentials
$\Delta_1$ and $\Delta_2$ in the two wires.  The Hamiltonian for
such a system reads
\begin{equation}
\hat{H}= \hat{H}_{SC} + \hat{H}_R, \label{H}
\end{equation}
where $\hat{H}_{SC}$ is Hamiltonian of the nanowire in the presence
of external magnetic fields and $\hat H_R$ represents Rashba SOI.
The former term is given by
\begin{eqnarray}
\hat{H}_{SC} &=& \int dx \Big\{  \sum_{\sigma, \sigma'}
\psi_{\sigma}^{\dag}(x) \Big([\xi_{\hat k}+U(x)] \sigma_0 + h
\sigma_z \nonumber\\
&& + B \{[\sigma_x \cos \phi_1 + \sigma_y \sin \phi_1] \theta(-x) +
[\sigma_x \cos \phi_2 \nonumber\\
&& + \sigma_y \sin \phi_2]
\theta(x) \} \Big) \psi_{\sigma'}(x) \nonumber\\
&& +  (\Delta_1 \theta(-x) + \Delta_2 \theta(x))
\psi_{\uparrow}^{\dag}(x) \psi_{\downarrow}^{\dag}(x)+ {\rm h.c.}
\Big\}, \label{H-sc}
\end{eqnarray}
where $\xi_{\hat k} = \epsilon
\left(\frac{\hbar}{i}\frac{\partial}{\partial x}\right) -\epsilon_F$
denotes the electron kinetic energy as measured from the Fermi
energy $\epsilon_F$, $\psi_{\sigma}(x)$ is the electron annihilation
operator, $h$ and ${\bf B}$ are external Zeeman magnetic fields in
$z$ direction and in the $x-y$ plane respectively, $\theta(x)$ is
the Heaviside step function, and $\sigma_{x,y,z}$ and $\sigma_0$
denote Pauli and identity matrices respectively in spin space. Note
that the magnetic field ${\bf B}$ forms an angle $\phi$ with wire
which can be tuned externally. In what follows, we choose ${\bf B}$
in the left side of the junction to be aligned along the wire
($\phi_1=0$) while in the right side it is chosen to make an angle
$\phi$ with it ($\phi_2= \phi$). In Eq.\ (\ref{H-sc}), the pairing
potential $\Delta_2$ in the right of the junction is chosen to have
a phase difference $\varphi$ compared to its left counterpart:
$\Delta_2 = |\Delta| \exp(i\varphi)$ and $\Delta_1 = |\Delta|$.  The
potential barrier $U(x)=U_0 \delta (x)$ represents the barrier
potential between two superconductors located at $x=0$. The
Hamiltonian of Rashba SOI can be written as
\begin{eqnarray}
\hat{H}_{R}= \sum_{\sigma, \sigma'} \int dx \psi_{\sigma}^{\dag}(x) \alpha \left[v_x \sigma_z\right]\psi_{\sigma'}(x),
\label{rashba-o}
\end{eqnarray}
where $\alpha$ is the strength of Rashba SOI which is chosen to be
the same for both wires. In what follows, we shall look for the
localized subgap Andreev bound states with $\epsilon(k)< |\Delta|$
for the Josephson junction of two nanowires described by Eq.\
(\ref{H}).

\section{\textbf{Andreev bound states, Josephson and magneto-Josephson effects}}
\label{sec3}

In this section, we first obtain solution for the Andreev bound
states for junction described by Eq.\ (\ref{H}). To do this, it is
advantageous to use a four component field operator given by
\begin{eqnarray}
\Psi^{\dag}_a(x)=\left(\psi_{a,\uparrow, +}^{\dag}(x), \psi_{a,
\downarrow,+}^{\dag}(x), \psi_{a, \downarrow, -} (x), \psi_{a,
\uparrow, -}(x)\right) \label{op1}
\end{eqnarray}
Here the third subscript of the annihilation operator (which we
shall designate henceforth as $b$) labels the right- ($b=+$) and the
left-moving $(b=-$) quasiparticles respectively while the index
$a=R,L$ denotes either right ($R=-$) or left ($L=+$) superconductor.
In terms of the field operator given by Eq.\ (\ref{op1}), the
Hamiltonian (Eq.\ (\ref{H})) can be written as $\hat{H}= \sum_{a=R,L}
\int dx \Psi^{\dag}_a(x) \mathcal{H}_a \Psi_a(x)$ using the Pauli
matrices $\sigma_i$ in spin- and $\tau_i$ in particle-hole spaces.
From Eqs. (\ref{H}) and (\ref{H-sc}), we find
\begin{eqnarray}
&& \mathcal{H}_R=\xi_{k, b} \tau_z \sigma_0 + h \tau_0 \sigma_z -i k
\alpha  \tau_z \sigma_z \label{H2} \\
&& + B \tau_z \left(\sigma_x \cos \phi + \sigma_y \sin \phi  \right)
+|\Delta| (\tau_x \cos \varphi - \tau_y \sin \varphi) \sigma_z,
\nonumber
\end{eqnarray}
and ${\mathcal H}_L= {\mathcal H}_R (\phi=0;\varphi=0)$. In Eq.\
(\ref{H2}), the energy spectrum of the electrons are linearized
around the positive and negative Fermi momenta leading to $\xi_{k,
b}= b v_F\left(-i \frac{\partial}{\partial x} - k_F\right)$, where
$v_F$ is the Fermi velocity. Note that the Hamiltonians $\mathcal
{H}_{R,L}$ acquires a magnetism-superconductivity duality
\cite{nab08, jpar13} in the absence of the kinetic term, implying
that it becomes invariant under the transformation $\{\Delta,
\epsilon_F, \varphi, \tau_i \} \to \{B, h, - \phi, \sigma_i \}$. The
existence of a magneto-Josephson effect in a topological insulator
is known to be a result of this duality \cite{jpar13}.
We shall see that for the system we study, the
magneto-Josephson effect takes place even in the presence of the
additional quadratic kinetic energy term of the electrons.

The energy spectrum of quasi-particles in a bulk superconductor in
the presence of SOI and external magnetic fields and its expression
for different asymptotic is calculated in Appendix\ \ref{appa}. Note
that in our case, all energies are measured from the Fermi energy;
thus the condition for realization of a topological superconducting
phase with effective $p$-wave pairing is $|\Delta|^2
> B^2+h^2$, \cite{jpar13}. However, the existence of such a topological phase
requires strong $B$ or $h$ and SO interaction so that only the
electron band of a single spin species remains below the Fermi
surface. In what follows we shall focus on the other regime where
the bands of both spin species are below the Fermi surface and the
superconductivity is still s-wave.

The Bogolyubov-de Gennes (BdG) equations for the superconductors in
the right- and left parts of the barrier are written as
\begin{equation}
\mathcal{H}_a \mathbf{\eta}_a(x)= E \mathbf{\eta}_a(x), \quad a= R,
L\label{Sch}
\end{equation}
where $\eta_a(x)$ denotes the BdG wave function. For a barrier
modeled by the delta function potential $U(x)= U_0 \delta (x)$, they
satisfy the boundary condition
\begin{eqnarray}
\mathbf{\eta}_L(0)= \mathbf{\eta}_R(0),\qquad \qquad
\partial_x \mathbf{\eta}_R -
\partial_x \mathbf{\eta}_L= k_F Z \mathbf{\eta} (0),
\label{bc}
\end{eqnarray}
where $Z=2mU_0/\hbar^2k_F$ and the transmission coefficient $D$ is
expressed through $Z$ as $D=4/(Z^2 +4)$.

The constructed wave functions $\eta_a(x)$ with the boundary conditions
\ref{bc} yield the energy of the Andreev bound states for our system.
We first note that the Rashba SOI
splits the energy spectrum shifting it along the momentum axis,
and results in four Fermi momenta at $k= \pm k_{F \pm}$ (see, Eqs.\
(\ref{E100}) and (\ref{k100})). The contribution to the Andreev
bound states comes from momenta around these Fermi points. The
external magnetic field splits spin-up and spin-down electrons (see,
Eqs.\ (\ref{E011}) and (\ref{k011})) even in the absence of SOI,
and the amplitudes of the electron wavefunction are redistributed
around four Fermi points due to the presence of such a field.
Finally, the presence of a barrier between the two superconductors
leads to  superposition of the right and left moving quasiparticles.
Therefore, the BdG wavefunction $\eta_a(x)$ can be written,
as it is shown in Appendix \ \ref{appa2} under (\ref{wave}),
as a linear superposition of its right and left moving components
with coefficients $A_a$, $B_a$, $C_a$, and $D_a$
around each Fermi momentum and with two different spins.

Substituting the wave functions (\ref{wave}) into the boundary
conditions (\ref{bc}) one gets eight linear homogeneous equations
for $A_a$, $B_a$, $C_a$, and $D_a$ with $a= \pm$ which can be
represented in terms of a $8 \times 8$ matrix $\Lambda$ and a column
vector $\Phi = ( A_a, B_a, C_a, D_a)^T$ as $\Lambda \Phi =0$. The
details of this procedure is charted out in Appendix \ \ref{appa2};
here, we simply note that, as shown in App.\ \ref{appa}, the
quantities $F_{\sigma \sigma'}$ which are determinants of selected
blocks of the matrix $\Lambda$ (Eqs.\ \ref{arrayeq1} and
\ref{arrayeq2}), plays a crucial role in these computations. The
energy of the Andreev bound states can then be obtained from ${\rm
Det} \Lambda =0$ and thus depend on $F_{\sigma \sigma'}$. We note
that since the momentum splitting $k_+-k_-$ vanishes in the absence
of SOI and magnetic field; in this limit, either $A_a +C_a \to A_a$
and $B_a + D_a \to B_a$ or both $C_a$ and $D_a$ vanish. The elements
of four columns of the $8 \times 8$ determinant, depending on $k_+$
become equal to other four column elements as $k_+ = k_-$, and the
determinant $\Lambda$ vanishes as $\alpha \to 0$ and $B,h \to 0$.


{\it Andreev bound states at $\alpha = |{\bf B}| = h=0$:} In this
limit, the Andreev bound states are determined using $4 \times 4$
determinant written for electron and hole pairs with opposite spins
\cite{ksy04}. The details of the calculation is given in \ref{appb}.
In order to get the explicit expressions for the wave functions
$\eta_{a,\sigma,b}$ and $\eta_{a,{\bar \sigma},b}^{\ast}$  we write
Eq.\ (\ref{Sch}) for finite $B$, $h$ and $\alpha$ as
\begin{eqnarray}
&& (E + i a b v_F k + iab \alpha k - h)\eta_{a, \uparrow, b} -
Be^{-i
\phi_a} \eta_{a, \downarrow, b}  \nonumber\\
&&- \Delta_a \eta^{\ast}_{a, \downarrow, \bar{b}}= 0 \label{Sch1}\\
&&(E + i a b v_F k - i ab \alpha k + h)\eta_{a, \downarrow, b} -
Be^{i
\phi_a} \eta_{a, \uparrow, b} \nonumber\\
&& + \Delta_a \eta^{\ast}_{a, \uparrow, \bar{b}} = 0 \label{Sch2}\\
&& (E - i a b v_F k - i ab \alpha k - h)\eta^{\ast}_{a, \downarrow,
\bar{b}}+ Be^{-i \phi_a} \eta^{\ast}_{a, \uparrow, \bar{b}}  \nonumber\\
&&- \Delta_a^{\ast} \eta_{a, \uparrow, b} = 0 \label{Sch3}\\
&& (E - i a b v_F k + i a b \alpha k + h)\eta^{\ast}_{a, \uparrow,
\bar{b}} + Be^{i \phi_a}
\eta^{\ast}_{a, \downarrow, \bar{b}} \nonumber\\
&&+\Delta_a^{\ast} \eta_{a, \downarrow, b}= 0. \label{Sch4}
\end{eqnarray}
We now use Eq.\ \ref{Sch4} to compute the Andreev bound state energy
at $\alpha = B= h=0$. The details of the calculation is charted out
in Appendix \ \ref{appb}. As shown in Appendix \ \ref{appb}, the
contribution to the bound state energy comes only from expression of
$F_{\sigma {\bar \sigma}}^{\ast}(k)$, and all other ratios vanish.
By equating $F_{\sigma {\bar \sigma}}^{\ast}(k)$ (Eq.\
(\ref{F-up-down})) to zero and using the expressions (\ref{E000})
and (\ref{k000}) for the energy and momentum in this limit, one gets
an expression for the bound state energy in consistent with the
well-known result \cite{abold,zag1,ksy04},
\begin{equation}
E_0= \pm |\Delta|\sqrt{1-D\sin^2\frac{\varphi}{2}}.
\label{E0}
\end{equation}
Thus our formalism reproduces the earlier known result in the
literature in this limit.

{\it Absence of Rashba SOI}: In this case, $\alpha =0$ and ${\bf B},
h \neq 0$, the main contribution, which depends on the magnetic
field orientation, yields the expression (\ref{cont2}) with
(\ref{energyM}) for $F_{\uparrow \uparrow}^{\dag}(k)$ and
$F_{\downarrow \downarrow}^{\dag}(k)$. Although the contribution
from (\ref{cont1}) does depend on the magnetic field, it does not
depend on the field orientation $\phi$. A few lines of algebra then
leads to the equation for the energy of the Andreev bound states,
obtained by equating the sum of (\ref{ksy}), (\ref{cont1}) and
(\ref{cont2}) to zero, using (\ref{E011}) and (\ref{k011}) for the
energy spectrum and momentum in this limit, given by
\begin{widetext}
\begin{eqnarray}
&& \left(|\Delta|^2- E^2_s -  D|\Delta|^2 \sin^2 \frac{\varphi}{2}\right)^2 +\frac{4E_s D |\Delta|^2 B^4 \sqrt{B^2+h^2}}
{(h+\sqrt{B^2+h^2})^4}\left(\sin^2 \frac{(\varphi - \phi)}{2}-\sin^2 \frac{(\varphi + \phi)}{2}\right)+\nonumber\\
&&16 h\sqrt{B^2+h^2} D \sin^2 \frac{\varphi}{2}\left(1-D\sin^2\frac{\varphi}{2}\right)
\left\{E^2_s-2\left(1-D\sin^2\frac{\varphi}{2}\right)\left[|\Delta|^2-4(h^2+B^2)\sin^2\frac{\varphi}{2}\right]\right\} =0,
\label{eq-mag-add}
\end{eqnarray}
where the second and third terms come from (\ref{cont2}) and
(\ref{cont1}) corresponding to the reflection mechanisms
(\ref{dagup-up}) and (\ref{dagup-down}). If we neglect the third
contribution, which can be done for $h \ll B$, Eq.\
(\ref{eq-mag-add}),can be written
\begin{eqnarray}
E^2_s - |\Delta|^2 + D|\Delta|^2 \sin^2 \frac{\varphi}{2} \pm
  |\Delta| \frac{2 B^2}{(h+\sqrt{B^2+h^2})^2}
\sqrt{E_s D (B^2+h^2)^{1/2} |\sin \phi|~| \sin \varphi|} =0.
\label{eq-mag}
\end{eqnarray}
\end{widetext}
We find that Eq.\ (\ref{eq-mag}) leads to the following features of
the Andreev bound states. First, $E_s$ decreases with increasing the
magnetic field. Second, Eq.\ \ref{E0} is correctly recovered as $B
\to 0$.

\begin{figure} [h!]
\centering
\includegraphics[height=50mm]{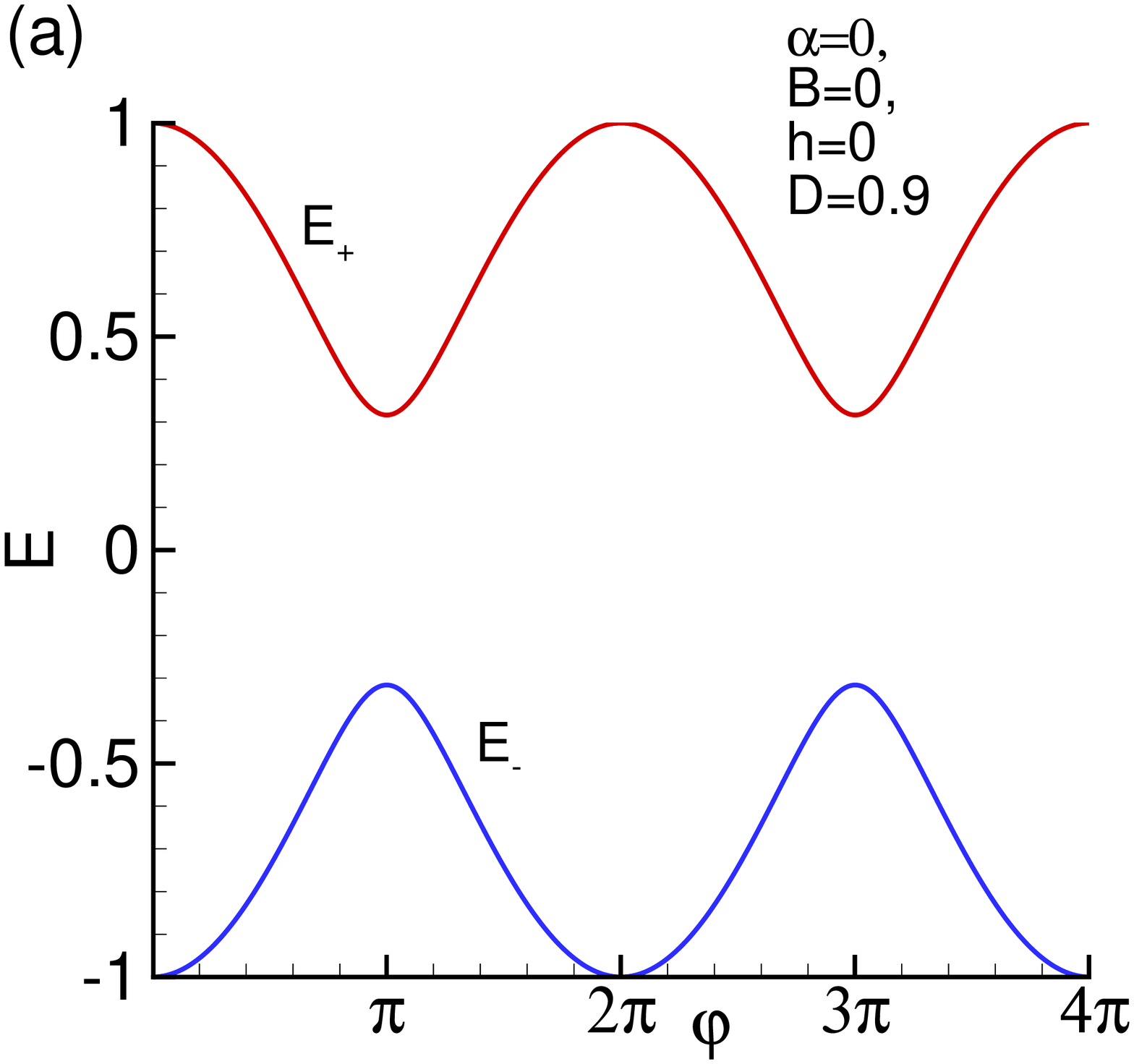}
\includegraphics[height=50mm]{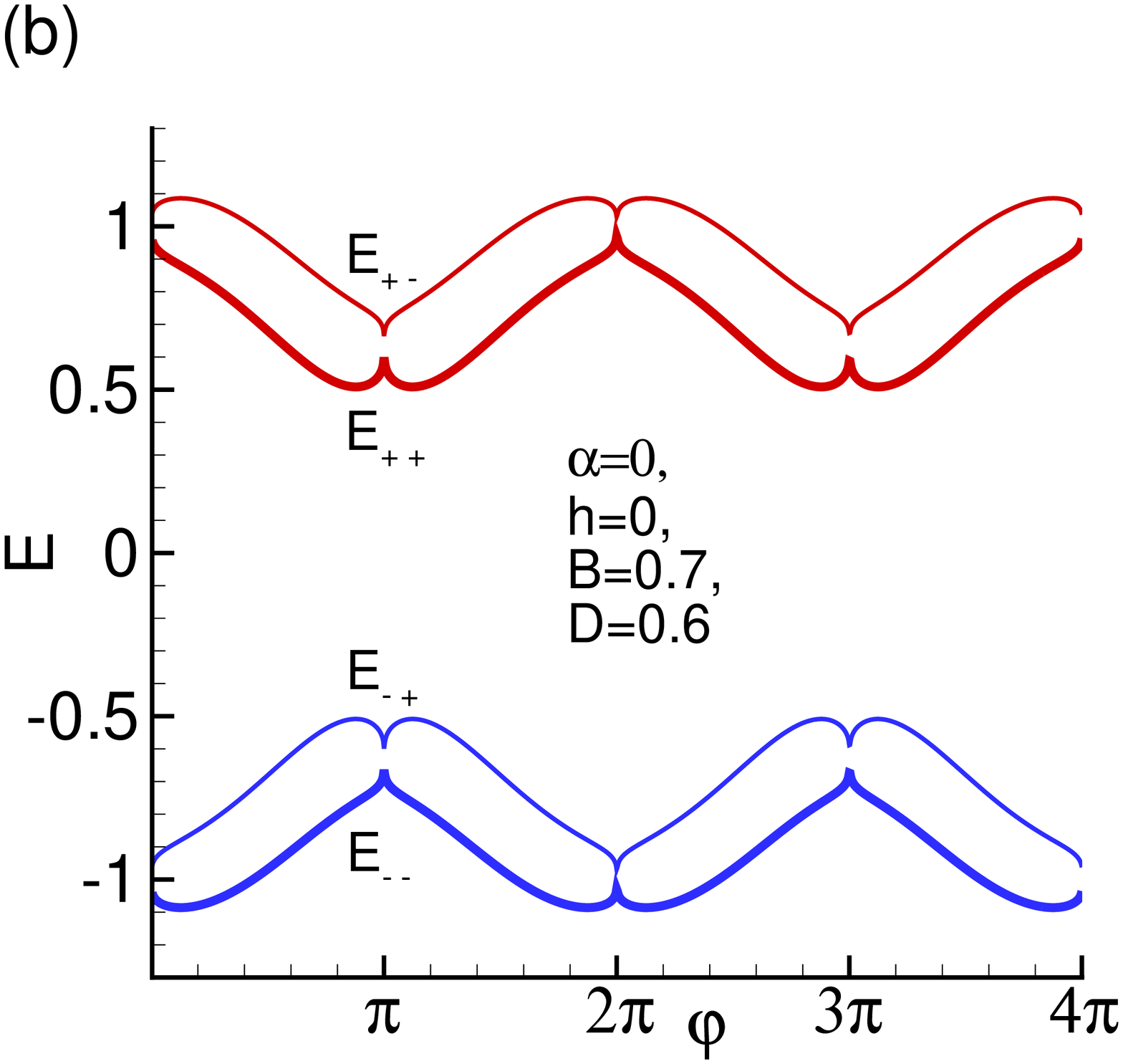}
\caption{Dependence of the branches energy, given by Eq.\ (\ref{eq-magA}), on
the order parameter phase difference at $\phi=0.5$ and (a) $B=h=0$,
$D=0.9$; (b)  $h=0.1$,  $B=0.9$, $D=0.3$.} \label{magnet}
\end{figure}

Eq.\ (\ref{eq-mag}) can be  solved approximately. We replace the
energy under square root by its zero-approximation value (\ref{E0}),
which yields $E_s(B,h) \equiv \pm E_s^M(B,h)$ with $s=\pm$,
where
\begin{widetext}
\begin{eqnarray}
E_s^M(B, h) = \Bigg\{ |\Delta|^2\left(1 - D \sin^2
\frac{\varphi}{2}\right) -s\frac{2 |\Delta|^{3/2}
B^2}{(h+\sqrt{B^2+h^2})^2} \sqrt{ D \sqrt{(B^2+h^2)
\left(1-D\sin^2\frac{\varphi}{2}\right)}~ |\sin \phi|| \sin
\varphi|}~\Bigg\}^{1/2}  \label{eq-magA}
\end{eqnarray}
\end{widetext}
The second term in the bracket of Eq.\ (\ref{eq-magA})depends on the
magnetic field as $\sim \sqrt{B}$ for $B \gg h$. We note here that
$E_s(B,h)$ oscillates both with the superconducting phase difference
$\varphi$ and the angle orientation $\phi$ of ${\bf B}$ with a
period $2 \pi$ as shown in Fig.\ \ref{magnet}. Note that all
parameters in the figures presented below are dimensionless ones in
the scale of $|\Delta|$, i.e. $ B \to B/|\Delta|$, $h \to
h/|\Delta|$, $E \to E/|\Delta|$. At $B=0$, Eq.\ \ref{E0} is
recovered for $s$-wave superconducting junction and the Andreev
bound state energy oscillates with $2\pi$ periodicity (see, Fig.\
\ref{magnet}a) for barrier transparency $D<1$. The electron-like and
hole-like energy branches corresponding to $\pm E_s^M(B,h)$, touch
each other at maximal transmission when $D=1$, creating a
zero-energy state at the center of the Brillouin zone. The variation
of $B$ and $h$ changes a character of $\varphi$- and
$\theta$-dependencies {\bf of $E_s^M$}. Note that since the gap
between them vanishes at $\varphi=\pi/2$, it might be possible to
have a $4 \pi$ periodic component of the Josephson current in case
of Landau-Zener transitions with a finite transmission probability
between two states. This case will be investigated somewhere else.


{\it Absence of in-plane Zeeman field}: Next, we consider the
Andreev bound states for $|{\bf B}|=0$, but $\alpha, h \neq 0$. We
find that Eqs. (\ref{Sch1})- (\ref{Sch4}) in this case link only
$\eta_{a,\sigma, b}$ and $\eta^{\ast}_{a, {\bar \sigma}, {\bar b}}$
and are hence greatly simplified. A few lines of algebra shows that
the Andreev bound states energy in this case can be expressed as
\begin{eqnarray}
\left[v_F^2 k^2-D \sin^2 \frac{\varphi}{2}\right]^2 + \nonumber \\
 F_{\uparrow,
\downarrow}^{\ast}(k_+) F_{\downarrow, \uparrow}^{\ast}(k_-)  -
F_{\uparrow, \downarrow}^{\ast}(k_-)F_{\downarrow,
\uparrow}^{\ast}(k_+)=0. \label{energyB=0}
\end{eqnarray}
The expression for $F_{\uparrow, \downarrow}^{\ast}(k)$ in this
limit is calculated in Appendix \ref{appb} and is  given by Eq.\
(\ref{ApF}). The expression for  $F_{\downarrow,
\uparrow}^{\ast}(k)$ at ${\bf B}=0$ is obtained from Eq.\
(\ref{ApF}) by replacing $\alpha \to - \alpha$ and $h \to -h$. Below
we will study two asymptotic solutions of Eq.\ (\ref{energyB=0}) at
$h=0$, $\alpha \neq 0$ and $\alpha = 0$, $h \neq 0$. In the former
case, Eq.\ (\ref{energyB=0}) with (\ref{ApF}) yields the following
equation

\begin{eqnarray}
\left(|\Delta|^2 -E^2- |\Delta|^2D \sin^2\frac{\varphi}{2}\right)^2-\nonumber\\
E^2\frac{16v_F^2 \alpha^2}{(v_F^2- \alpha^2)^2}
\left(|\Delta|^2 -E^2- |\Delta|^2D \sin^2\frac{\varphi}{2}\right)D\sin^2\frac{\varphi}{2} -\nonumber\\
E^4\frac{16v_F^2 \alpha^2}{(v_F^2- \alpha^2)^2} D^2 \sin^4\frac{\varphi}{2}=0.\hspace{0.5cm}
\end{eqnarray}

Solution of this equation provides a simple expression for the Josephson energy
\begin{equation}
E_s=\pm |\Delta|\left\{\frac{1-D\sin^2\frac{\varphi}{2}}{1 -  \frac{s~4v_F \alpha}{(v_F + s~ \alpha)^2}
D \sin^2\frac{\varphi}{2}}\right\}^{1/2} \equiv \pm E_s^{SOI},
\label{SOI}
\end{equation}
where the sign $\pm$ in the front of the expression signifies an
electron and hole energies, whereas the sign $s= \pm$ characterizes
Rashba splitting of the electron and hole states. This expression
shows that $E_{s}$ depends nonlinearly on the SOI coupling constant
$\alpha$. We note that Eq.\ \ref{E0} is once again recovered as
$\alpha \to 0$. According to (\ref{SOI}),  $E$ oscillates still with
$2\pi$ period for $D < 1$ and  $\tilde{\alpha}=\alpha/v_F \ll 1$,
which is presented in Fig.\ \ref{solutions}(a) at $\alpha=0.1$ and
$D=0.6$. Possible solutions for the energy spectrum according to the
expression (\ref{SOI}) as a function of the order parameter phase
difference at $\alpha=0.205$ and $D=1$ is presented in Fig.\
\ref{solutions}(b). It shows touching of all four branches at
$\varphi = \pi$.  The electron- and hole energy branches approach
each other faster for non-zero SOI.

The dependence of the $E_{+-}^{SOI}$ energy branches on the order
parameter phase difference $\varphi$ at fixed transmission
coefficient $D$ and different values of the SOI strength $\alpha$,
is presented in the left panel of Fig.\ \ref{e-p-dep}(a). In Fig.\
\ref{e-p-dep}(b), we present the dependence of the Andreev bound
state energies on $D$ for fixed $\alpha$. We note that both the
branches approach zero as $\alpha$ or $D$ is varied.

\begin{widetext}
{\it Absence of $B$ and $\alpha$}: Next, we consider the case $|{\bf
B}|,\alpha = 0$ but $h \ne 0$. In this case, Eq.\ (\ref{energyB=0})
reduces to

\begin{figure} [h!]
\centering
\includegraphics[height=60mm]{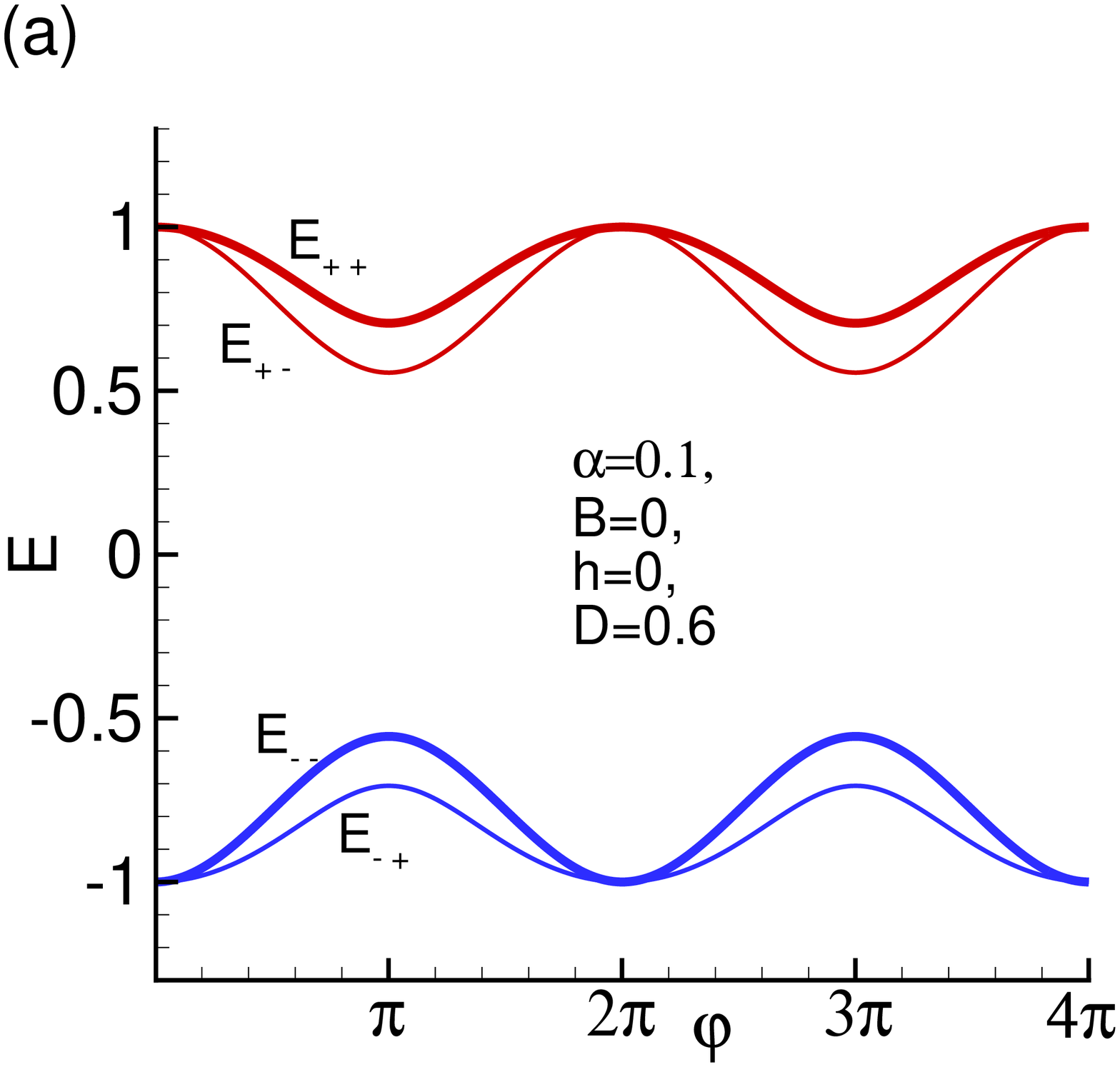}\includegraphics[height=60mm]{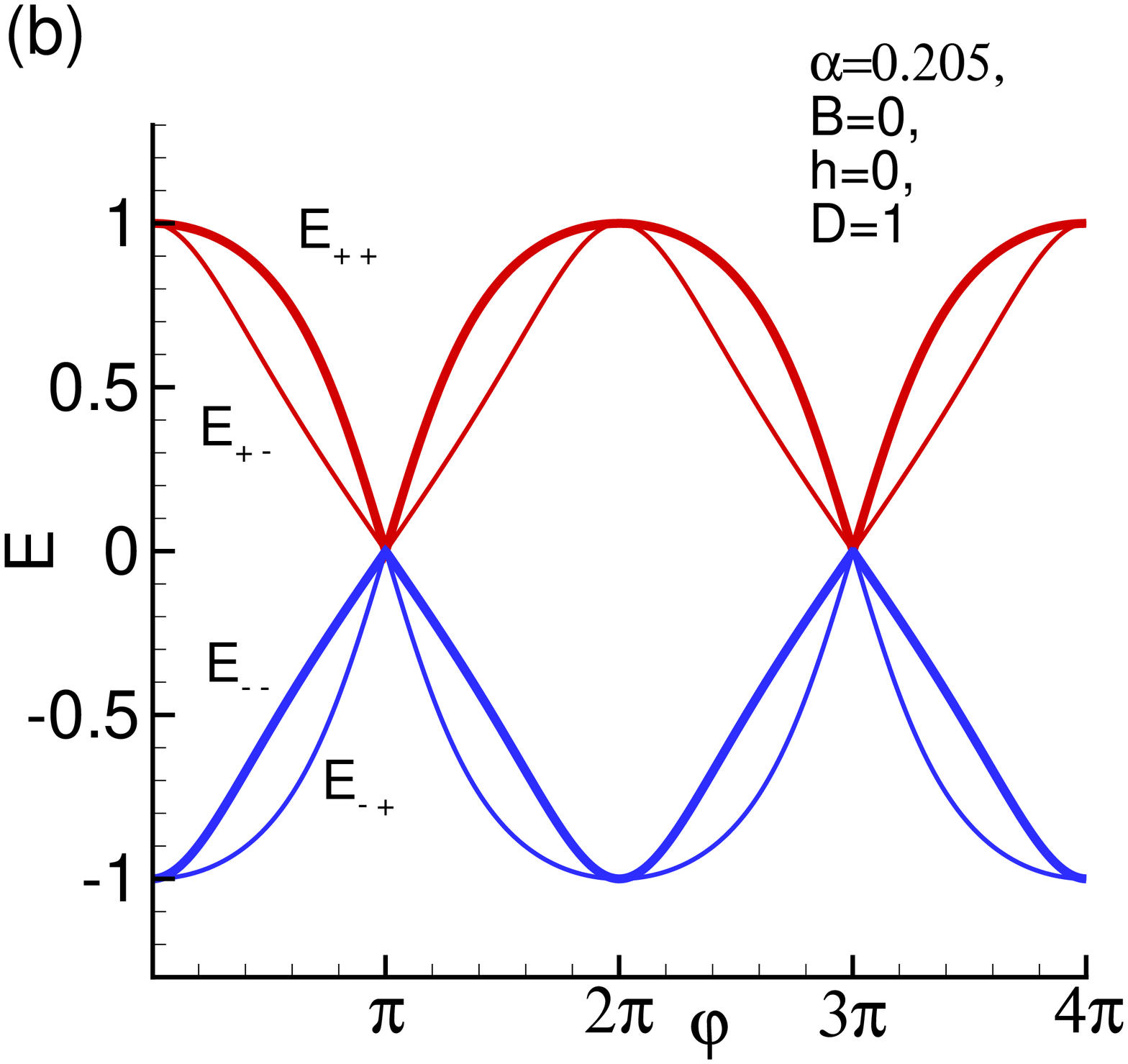}
\caption{Possible solutions for the branch's energy spectrum according to the
expression (\ref{SOI}) as a function of the order parameter phase
difference at (a) $\alpha=0.1$ and $D=0.6$, and (b) $\alpha=0.205$
and $D=1$.} \label{solutions}
\end{figure}

\begin{figure} [h!]
\centering
\includegraphics[height=60mm]{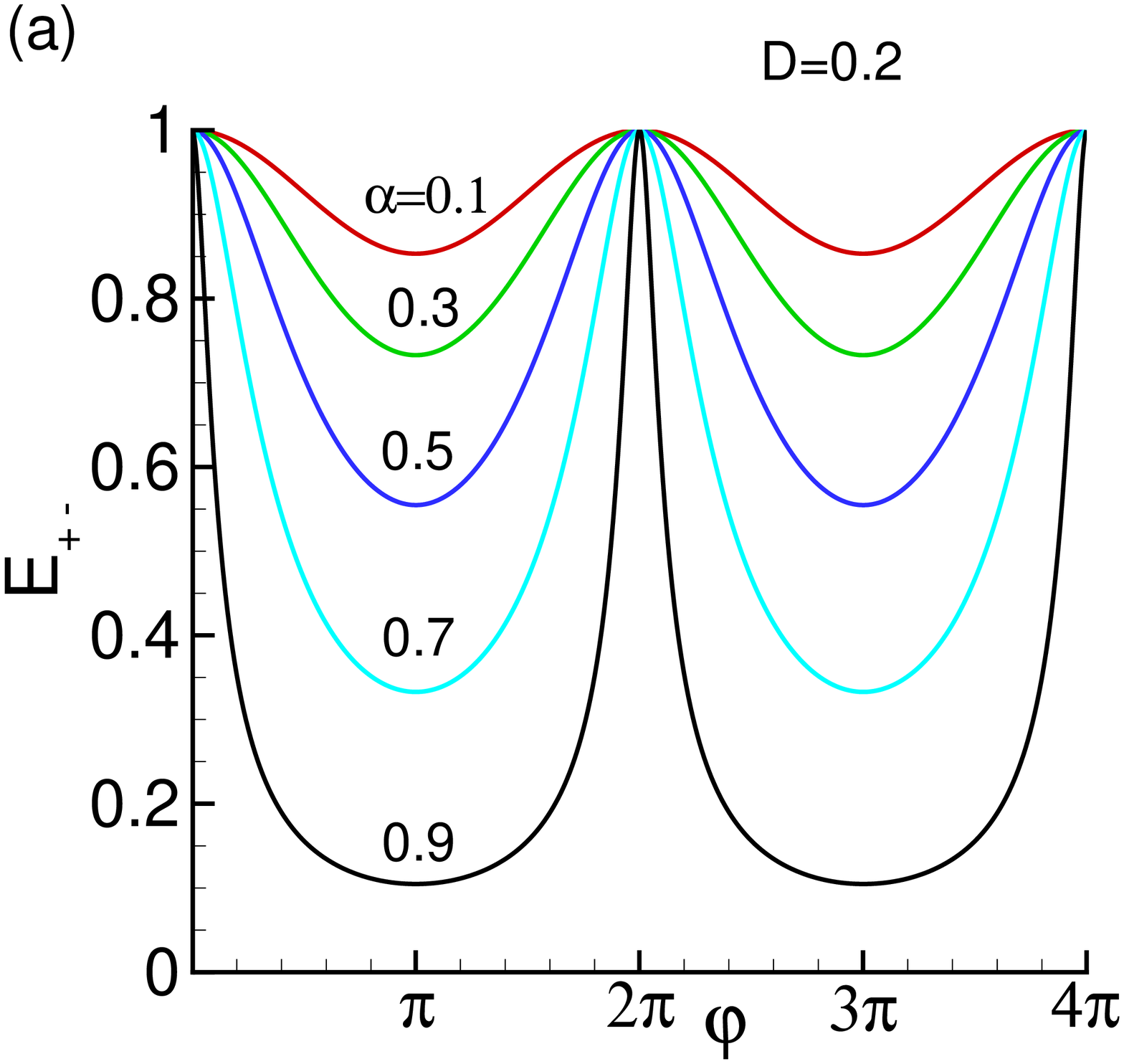}\includegraphics[height=60mm]{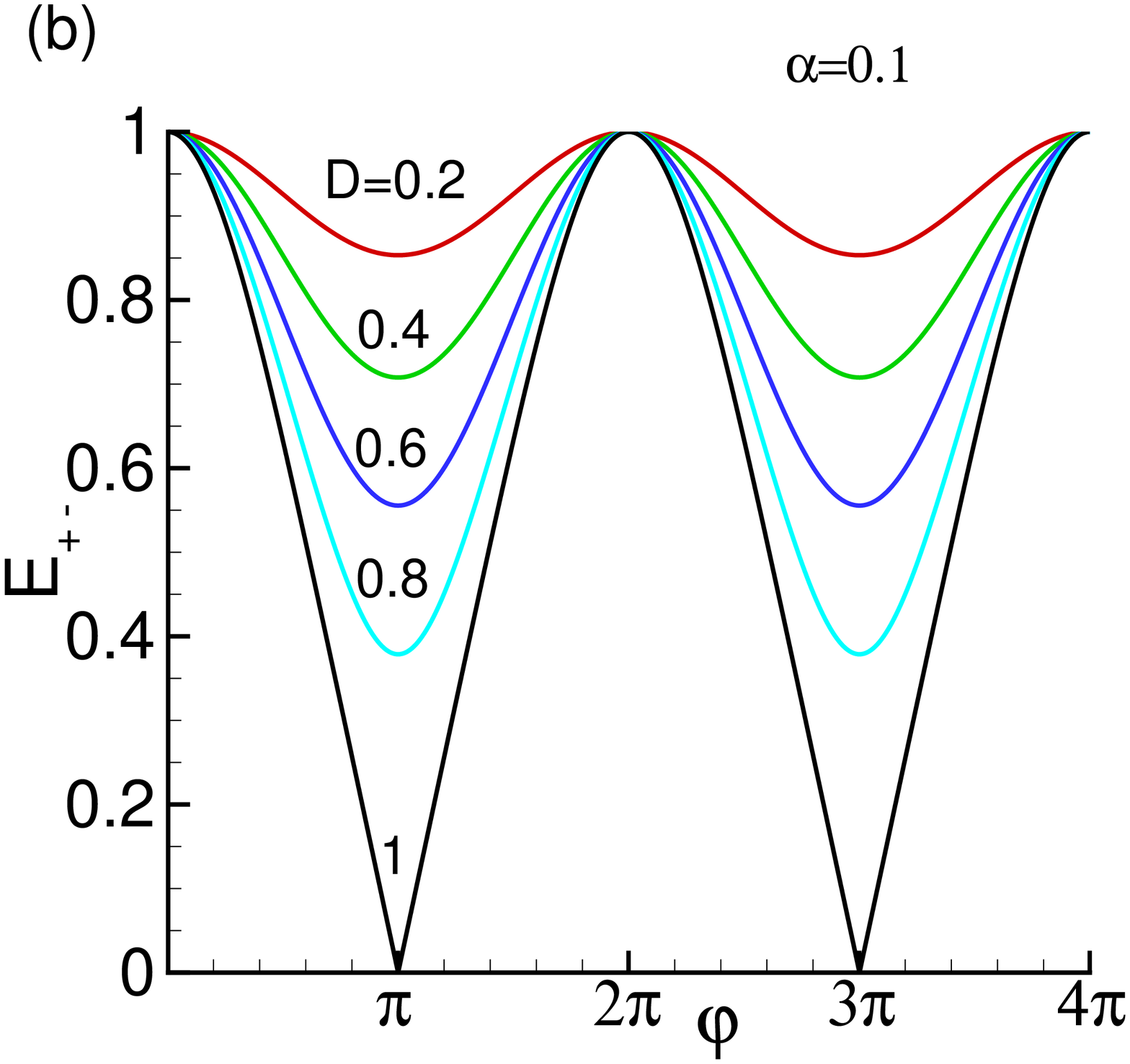}
\caption {Dependence of the branch's energy on the
order parameter phases difference at (a) $D=0.2$ and different
values of the SOI strength, and  (b) $\alpha=0.1$ and different
values of the transmission coefficient $D$.} \label{e-p-dep}
\end{figure}

\begin{figure} [h!]
\centering
\includegraphics[height=45mm]{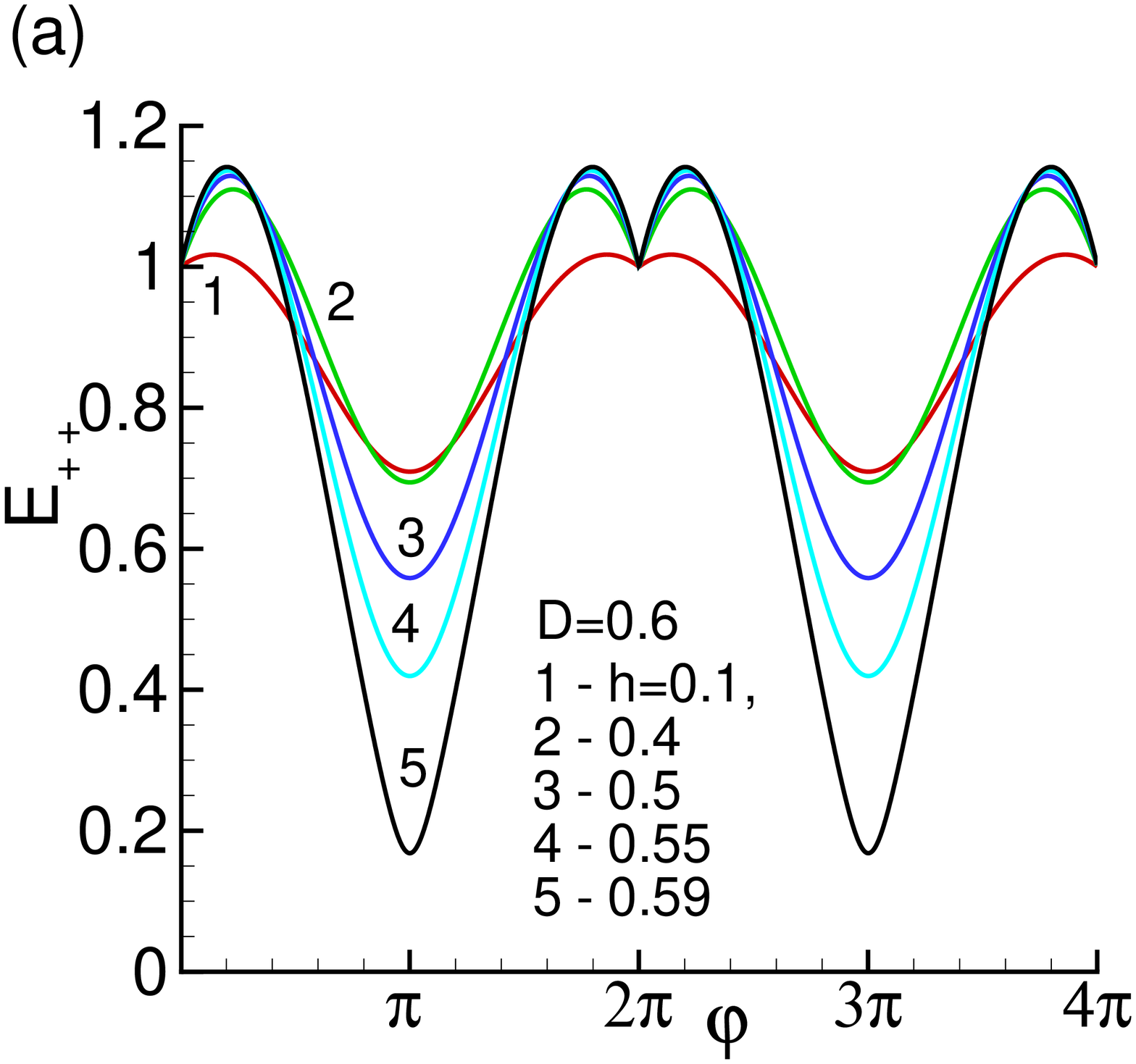}\includegraphics[height=45mm]{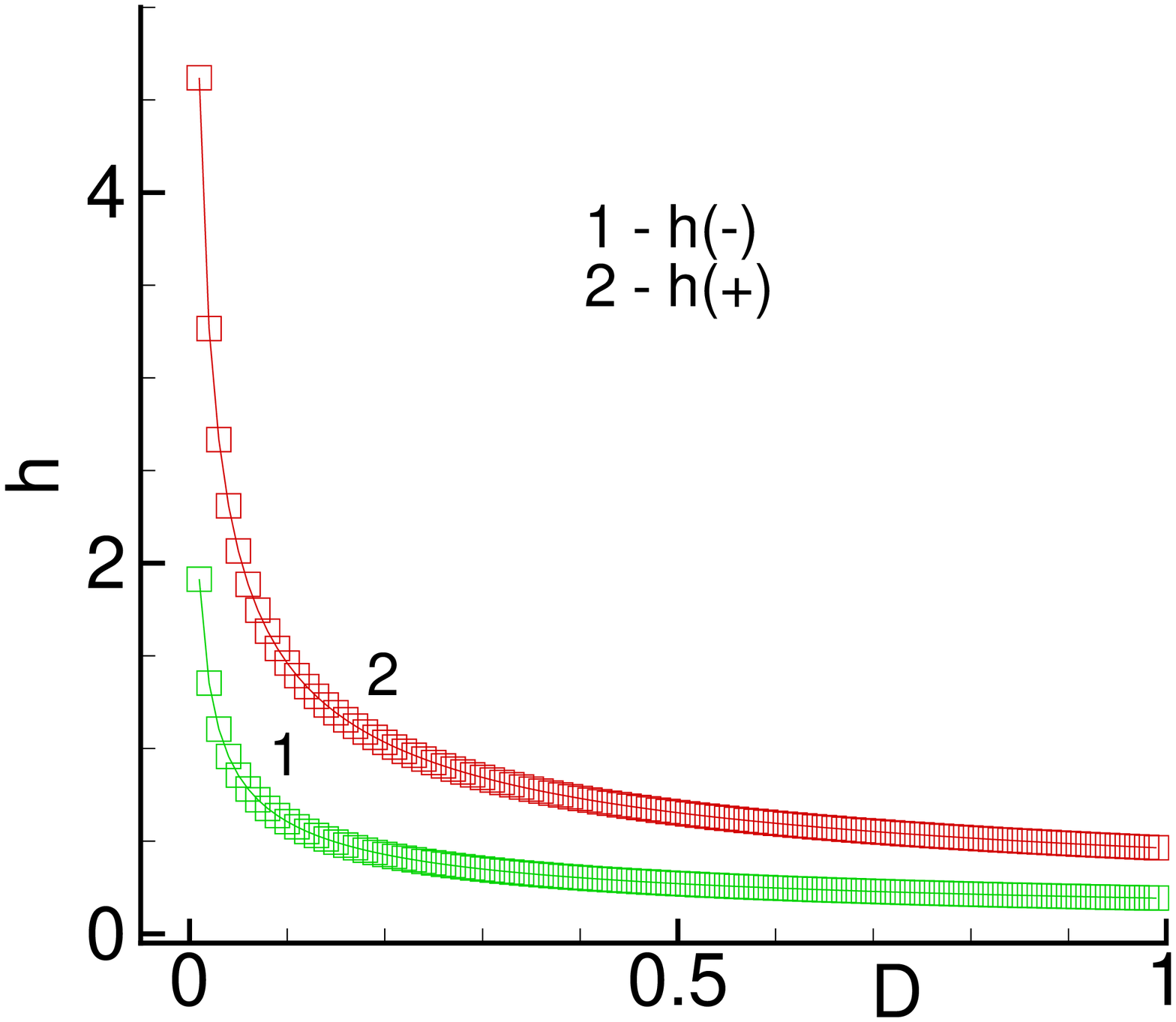}\includegraphics[height=45mm]{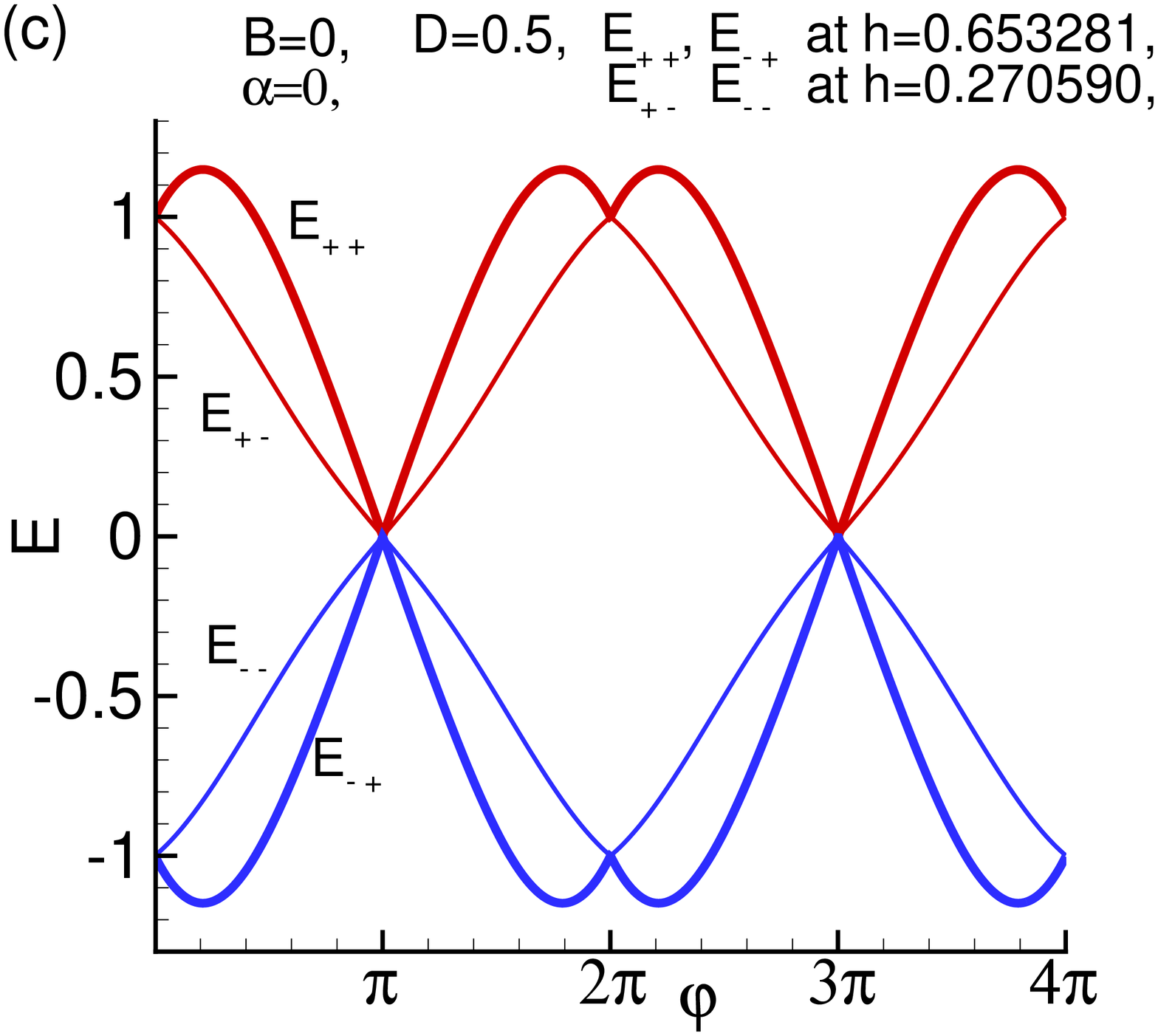}
\caption{(a) Dependence of the spin-up particle branch's energy  on
$\varphi$ at different $\tilde{h}$. Amplitude of the energy
oscillation increases with $\tilde{h}$; (b) Mutual optimal values of
$\tilde{h}$ and $D$ at which electron- and hole-energy branches are
crossed, creating a zero-energy mode; (c) The Andreev bound
state energies $E_{h \pm}$ touches at particular values of $D=0.5$
and $\tilde{h}=0.653281$ (thin curves), and of $D=0.5$ and
$\tilde{h}=0.270590$ (thick curves) which may make the oscillation
period $4 \pi$ in the Landau-Zenner sense.} \label{crossing}
\end{figure}

\begin{eqnarray}
&&\left(|\Delta|^2 -E^2_h-D|\Delta|^2 \sin^2 \frac{\varphi}{2}\right)^2 + 16 E^2_h h^2 D \sin^2\frac{\varphi}{2}
\left(1-D\sin^2\frac{\varphi}{2}\right) - \nonumber \\
&&32h^2D \sin^2\frac{\varphi}{2}\left(|\Delta|^2-4h^2D \sin^2\frac{\varphi}{2}\right)
\left(1-D \sin^2\frac{\varphi}{2}\right)^2 =0,
\end{eqnarray}
whose solutions read
\begin{eqnarray}
E_h(h)=\pm \sqrt{1-D\sin^2\frac{\varphi}{2}}\left\{|\Delta|^2 - 8h^2
D\sin^2\frac{\varphi}{2} + s 4h \sqrt{D}
\left|\sin\frac{\varphi}{2}\right|\sqrt{|\Delta|^2 - 4 h^2
D\sin^2\frac{\varphi}{2}} \right\}^{1/2}, \label{h}
\end{eqnarray}
\end{widetext}
where $s= \pm$. We note that the particle-like and the hole-like
branches touch at zero energy; in order to investigate the possible
existence of a zero energy mode, which may create a $4\pi$
oscillatory component of the Josephson current in the Landau-Zenner
sense, we introduce a dimensionless magnetic field
$\tilde{h}=h/|\Delta|$. It is easy to see from Eq.\ (\ref{h}) that the
condition for the particle and the hole states to cross at a phase
difference $\varphi$ is given by
\begin{eqnarray}
\left(\sqrt{1 - 4 \tilde{h}^2 D\sin^2\frac{\varphi}{2}} + 2s\tilde{h}\sqrt{D}\left|\sin\frac{\varphi}{2}\right|\right)^2- \nonumber \\
8\tilde{h}^2 D \sin^2\frac{\varphi}{2}=0,
\end{eqnarray}
which yields
\begin{equation}
\tilde{h}_s^2=\frac{1}{4 D[1+(\sqrt{2}-s)^2]\sin^2\frac{\varphi}{2}}.
\end{equation}

For $\varphi=\pi$, the value of the critical ${\tilde h}$ for
spin-up ($s=+1$) and spin-down ($s=-1$) states are ${\tilde
h}_+=1/\sqrt{4.6863 D}$ and ${\tilde h}_-=1/\sqrt{27.314 D}$,
correspondingly. We note here that the bands touch each other at $h
= \tilde h_s$ but do not cross; thus the Andreev states still have
$2 \pi $ periodic dispersion.

The variation of $E_h(h)$ with $\tilde{h}$, the dependence of
$\tilde{h}$ on $D$, the touching of the $E_{++}$ and $E_{-+}$ energy
branches at $D=0.5$ and $\tilde{h}=0.653281$, and that between
$E_{+-}$ and $E_{--}$ energy branches at D=0.5 and
$\tilde{h}=0.270590$ are plotted in Fig.\ \ref{crossing}. In Fig.\
\ref{crossing}(a), where  the dependence of the spin-up particle
energy branch on $\varphi$ at different $\tilde{h}$ is presented, we
find that the amplitude of the energy oscillation increases with
$\tilde{h}$, and additionally, the character of dependence around
$\varphi=2\pi$ is changed. In Fig.\ \ref{crossing}(b) we show the
mutual optimal values of $\tilde{h}$ and $D$ at which electron-like
and hole-like energy branches touche each other. Finally, the
touching of the two branches $E_{++}(h)$ and $E_{-+}(h)$ for $D=0.5$
and $\tilde{h}=0.653281$ is presented in Fig.\ \ref{crossing}(c). As
it was mentioned above, these feature might be responsible for a $4
\pi$ periodicity in case of Landau-Zener transitions with a finite
transmission probability between two states.

\section{Equilibrium Josephson current and spin current}
\label{sec4}

\begin{figure}
 \centering
\includegraphics[height=60mm]{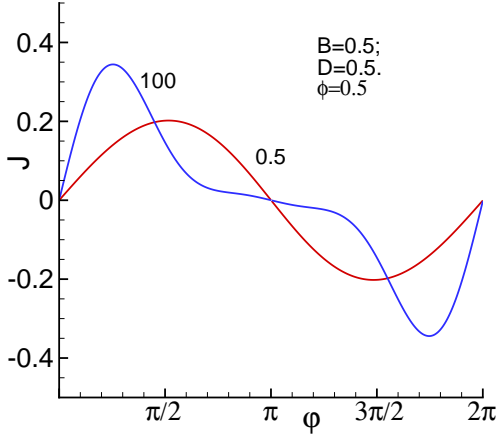}
\caption{Current-phase relation at magnetic field $h=0.1$ according to the formulas (\ref{currentT})-(\ref{cph})}
\label{44}
\end{figure}

The contribution of the Andreev bound state to the Josephson current
can be calculated using to the expression
\begin{equation}
J=\frac{2e}{h}\sum_{n} \frac{\partial E_n}{\partial \varphi}~f(E_n),
\label{current}
\end{equation}
where $n$ signifies all states which give a contribution to the
current, and $f(E_n)$ is the Fermi occupation number corresponding
to the $n$-th states. We note that since only the Andreev bound
states depend explicitly on the phase difference $\varphi$, their
expression can be used to determine the DC Josephson current using
Eq.\ (\ref{current}).  In the absence of SOI a contribution to the
total equilibrium current gives electron and hole states, each of
which is split into two levels due to Zeeman effect
\begin{equation}
J= - \frac{2e}{h}\sum_{s= \pm}\frac{\partial E_s^M}{\partial \varphi}\tanh\left(\frac{E_s^M}{2k_BT}\right),
\label{currentT}
\end{equation}
where the expression for $E_s^M$ is given by Eq.\ (\ref{eq-magA}),
and
\begin{widetext}
\begin{eqnarray}
\frac{\partial E_s^M}{\partial
\varphi}=\frac{1}{2E_s^M}\left[-\frac{D}{2}|\Delta|^2 \sin \varphi-
s \frac{B^2 |\Delta|^{3/2}\sqrt{D\sqrt{B^2+h^2}~|\sin
\phi|}~C(\varphi)} {4 (1-D \sin^2
\varphi/2)^{3/4}(h+\sqrt{B^2+h^2})^2 \sqrt{|\sin \varphi|}}\right]
\label{dEs}
\end{eqnarray}
with
\begin{equation}
C(\varphi)=\left\{
\begin{array}{rl}
-D \sin^2\varphi  +4\left(1-D\sin^2\frac{\varphi}{2}\right)\cos \varphi & \text{for} \quad 0 \le \varphi < \pi\\
D \sin^2\varphi -4\left(1-D\sin^2\frac{\varphi}{2}\right)\cos \varphi & \text{for} \quad \pi \le \varphi < 2\pi
\end{array}\right. \label{cph}
\end{equation}
\end{widetext}
The current-phase relation at magnetic field
$h=0.1$ calculated by using expressions (\ref{currentT}), (\ref{dEs}) and
(\ref{cph}) is presented in  Fig.\ \ref{44}. We note, that changes in $h$ does not make an essential effect at $h\leq B$.

Next, we consider the spin-Josephson current which is generated as
response to rotation of the magnetic field ${\bf B}=\{B \cos \phi,~B
\sin \phi,~0\}$ in $\{x,~y\}$ plane\cite{jpar13}. As shown in Ref.\
\onlinecite{jpar13}, the spin current can be defined as a derivative
of the tunneling energy with respect to the magnetic field
orientation $\phi$ and is given by
\begin{equation}
J_{MJ}=\sum_{s=\pm}\frac{\partial E_s^M}{\partial \phi}
\tanh\left(\frac{E_s^M}{2k_BT}\right), \label{J-mag}
\end{equation}

\begin{figure}
 \centering
\includegraphics[height=60mm]{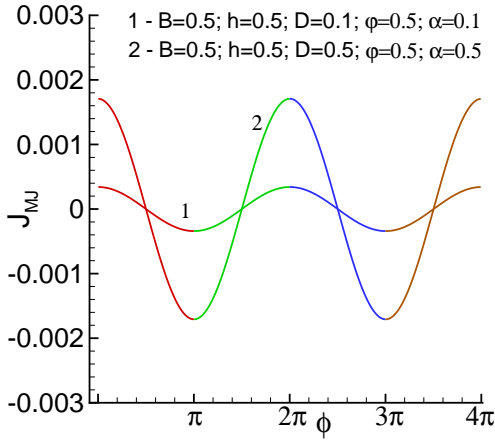}
\caption{Spin current as a function of magnetic field orientation at
$D=0.3$, $h=0.1$, $\alpha=0$, $\varphi=0.5$ and two values of magnetic filed $B=0.9$ (curve 1) and $B=2$ (curve 2).  Calculations are done according to the formulas
(\ref{J-mag}), (\ref{J-mag2}) and (\ref{eq-magA}).} \label{47}
\end{figure}

where
\begin{widetext}
\begin{equation}
\frac{\partial E_s^M}{\partial \phi}=-sl\frac{B^2
|\Delta|^{3/2}\sqrt{D \sqrt{(B^2+h^2)(1-D\sin^2 \varphi/2)}~|\sin
\varphi|}~\cos \phi}{2 E_s^M (h+\sqrt{B^2+h^2})^2 \sqrt{|\sin
\phi|}}, \label{J-mag2}
\end{equation}
\end{widetext}
with $l=1$ for $0 \le \phi < \pi$ and $l=-1$ for $\pi \le \phi
<2\pi$. As it is seen from formulas (\ref{dEs}) and (\ref{J-mag2}),
the product $E_s^M \frac{\partial E_s^M}{\partial \varphi}$
increases with $B$ at $B \gg h$ as $\sqrt{B}$. Instead in the
opposite limit when $B \ll h$ this product decreases with increasing
$h$ as $\sim B^2/h^{3/2}$. On the other hand, in the high
temperature limit, when $2k_BT \gg E_s^M$, one can expand $\tan x$
function for small argument $x \ll 1$ as $\tan x \sim  x$.
Therefore, the amplitude of the supercurrent $J$, given by Eq.\
(\ref{dEs}), and of the spin current $J_{MJ}$, given by Eq.\
(\ref{J-mag2}), will depend on the magnetic field exactly in the
same form as described above for two limiting cases. The change of
$h-$direction can rotate the direction of spin current.  Spin
current as a function of magnetic field orientation at two values of
magnetic filed $B=0.9$ and $B=2$   is shown in Fig.\ \ref{47}.
Calculations are done according to the formulas (\ref{J-mag}),
(\ref{J-mag2}) and (\ref{eq-magA}).

The Josephson current in other limiting case when $B=h=0$ and
$\alpha \neq 0$ is calculated by replacing $E_s^M$ with $E_s^{SOI}$
given by (\ref{SOI}) in the expression (\ref{currentT})
\begin{widetext}
\begin{equation}
J=\frac{e|\Delta|}{2h}\sum_{s = \pm}\frac{D \left(1-s\frac{4v_F \alpha}{(v_F+s \alpha)^2}\right)~
\sin \varphi}{\left[1-s\frac{4v_F \alpha}{(v_F +s\alpha)^2}D\sin^2\frac{\varphi}{2}\right]
\sqrt{\left(1-D \sin^2\frac{\varphi}{2}\right)\left[1-s\frac{4v_F\alpha}{(v_F+s\alpha)^2}D\sin^2\frac{\varphi}{2}\right]}}
\tanh\left(\frac{E_s^{SOI}}{2k_BT}\right).
\end{equation}
\end{widetext}

\begin{figure}[h!]
 \centering
\includegraphics[height=60mm]{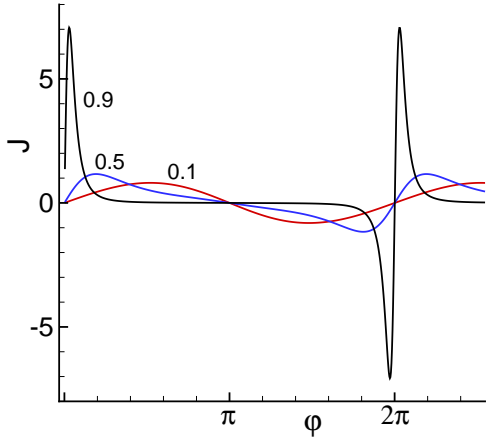}
\caption{ Transformation of current-phase relation with parameter of
spin-orbital coupling $\alpha$ at $D=0.5$ (formula (29)). Numbers show
the values of parameter spin-orbital coupling.} \label{49}
\end{figure}

The corresponding plots demonstrated a strong variation of
current-phase relation with parameter of spin-orbital coupling
$\alpha$ are presented in Fig.\ \ref{49}. The figure demonstrates a
crucial breaking of the sinusoidal current-phase relation  with
increase in spin-orbital coupling. It shows a singular behavior at
small $\varphi$.

\section{AC Josephson Effect}
\label{sec5}

In this section, we compute the AC Josephson effect for the tunnel
junctions mentioned above. If there is the voltage in Josephson junction
$V(t)=V_0+A\cos{\omega t}$, then from Josephson relation $\dot{\varphi}=2eV/\hbar$ we get

\begin{eqnarray}
\varphi(t) &=&  (2e/\hbar) [\varphi_0 + V_0t + \frac{A}{\omega}
\sin{\omega t}], \label{phaseeq1}
\end{eqnarray}

We shall now use this relation to obtain the Shapiro step width
for $B=h=0$ and demonstrate that the step-width depends on the
strength of the spin-orbit coupling. To do this we first consider
the case $\alpha=0$ for which $I_J[\phi]$ is given at $T=0$ by
\begin{eqnarray}
I_S = \frac{e\Delta}{4\hbar} \frac{D \sin \varphi(t)}{\sqrt{1 - (D/2) (1
- \cos \varphi(t))/2}} \label{iexp}
\end{eqnarray}
Substituting Eq.\ (\ref{phaseeq1}) into Eq.\ (\ref{iexp}), one gets
\begin{equation}
\ I_S = \frac{e\Delta}{2\hbar} \frac{D \sin({\varphi_0 + 2e
V_0t/\hbar + \frac{A}{\omega} \sin{\omega t}})}{\sqrt{1 - D (1 -
\cos({\varphi_0 + 2e V_0t/\hbar + \frac{A}{\omega} \sin{\omega
t}}))/2}} \label{ieq2}
\end{equation}
Using the identity
\begin{equation}
\Im \ e^{i(\varphi_0 + 2e V_0t/\hbar + \frac{A}{\omega} \sin{\omega t})} \\
=\Im \sum_n J_n \left(\frac{A}{\omega}\right) e^{i(\varphi_0 + t(2e
V_0/\hbar + n \omega))}
\end{equation}
where $\Im$ means imaginary part,  $n$ is an integer and $J_n$ denotes Bessel function of the
first kind, one gets
\begin{equation}
I_S = \frac{e\Delta}{2\hbar} \frac{D \Im \sum_n
J_n\left(\frac{A}{\omega}\right) e^{i(\varphi_0 + t(2e V_0/\hbar + n
\omega))})}{\sqrt{1 - D [1 - \Re{\sum_n J_n
\left(\frac{A}{\omega}\right) e^{i(\varphi_0 + t(2e V_0/\hbar + n
\omega))}}]/2}} \label{ieq3}
\end{equation}
Here $\Re$ means the real part.

The Shapiro steps thus occur when  $2e V_0/\hbar = -n_0 \omega$ for
integer $n_0$; at these values of the applied radiation frequency,
the AC component of the supercurrent vanishes leading to an extra
contribution to the dc current in the circuit. The magnitude of the
extra DC current from $I_s$ can be read off from Eq.\ (\ref{ieq3})
as
\begin{equation}
I_S^{\rm DC} = \frac{e\Delta}{2\hbar} \frac{D \sin \varphi_0 J_{n_0}
\left(\frac{A}{\omega}\right)}{\sqrt{1 - D [1 -
J_{n_0}\left(\frac{A}{\omega}\right) \cos(\varphi_0)]/2}}
\label{idc1}
\end{equation}

From Eq.\ (\ref{idc1}), we find that both the Shapiro step width and
the position of maxima/minima of $I_S$ depends on $D$. Let us assume
that the maxima and minima occur at $\pm \varphi_0^{n_0}(\omega)$.
Note that $\varphi_0^{n_0}(\omega)$ can be obtained from the
solution of $\partial I_S/\partial \phi_0 =0$ and equals $\pm \pi/2$
for $D \ll 1$. In terms of $\varphi_0^{n_0}(\omega) \equiv
\varphi_0^n$, one obtains the step width as
\begin{equation}
\ \Delta I_S = I_S^{\rm max} - I_S^{\rm min} = \\
\frac{e\Delta}{\hbar} \frac{D \sin (\varphi_0 ^n) J_{n0}
\left(\frac{A}{\omega}\right)}{\sqrt{1 - D [1 -
J_{n_0}\left(\frac{A}{\omega}\right) \cos(\varphi_0^n)]/2}}
\label{sstep1}
\end{equation}
which clearly shows the $D$ dependence of the step-width.

One can now carry out a similar analysis for the case where $B=h=0$
and $\alpha \ne 0$ (Eq.\ (\ref{SOI})). Starting from Eq.\
(\ref{current}), the AC Josephson current at $T=0$ can be obtained
as

\begin{widetext}
\begin{equation}
I_S^{\alpha} = \frac{e\Delta}{4\hbar} \\
\sum_{s=\pm} \frac{
D(1-\eta_s) \sin \varphi(t)}{\left[1-\eta_s D (1-\cos
\varphi(t))/2\right]^{3/2}\left[1-D (1-\cos
\varphi(t))/2\right]^{1/2}} \label{acjos1}
\end{equation}

where $\varphi(t) = \varphi_0 + 2e V_0t/\hbar + \frac{A}{\omega}
\sin{\omega t}$ and $\eta_s= 4 s \alpha v_F/(v_F + \alpha s)^2$.
Similar straightforward algebra, as carried out earlier in this
section, leads to steps at $n_0 \omega = - 2 e V_0/\hbar$ with

\begin{equation}
I_{\rm DC}^{\alpha} = \frac{e\Delta D J_{n_0}(\omega)
\sin\varphi_0}{4\hbar \left[1-D (1-J_{n_0}(\omega)\cos
\varphi_0)/2\right]^{1/2}}
\sum_{s=\pm}
\frac{(1-\eta_s)}{\left[1-\eta_s D (1-J_{n_0}(\omega) \cos
\varphi_0)/2\right]^{3/2}} \label{dcjos1}
\end{equation}

As before, the minimum and maximum of the DC component of the occurs
at $\pm \varphi_0^{n_0 \alpha}(\omega)$ which can be obtained as the
solution of $\partial I_{DC}^{\alpha}/\partial \varphi =0$. The step
width can thus be expressed in terms of $\varphi_0^{n_0
\alpha}(\omega) \equiv \varphi_0^{n_0 \alpha}$ as

\begin{equation}
\Delta I^{\alpha} = \frac{e\Delta D J_{n_0}(\omega)
\sin\varphi_0^{n_0 \alpha}}{2\hbar \left[1-D (1-J_{n_0}(\omega)\cos
\varphi_0^{n_0 \alpha})/2\right]^{1/2}}  \\
 \sum_{s=\pm}
\frac{(1-\eta_s)}{\left[1-\eta_s D (1-J_{n_0}(\omega) \cos
\varphi_0^{n_0 \alpha})/2\right]^{3/2}}\label{swidthso}
\end{equation}
\end{widetext}

\begin{figure}[h!]
 \centering
\includegraphics[height=50mm]{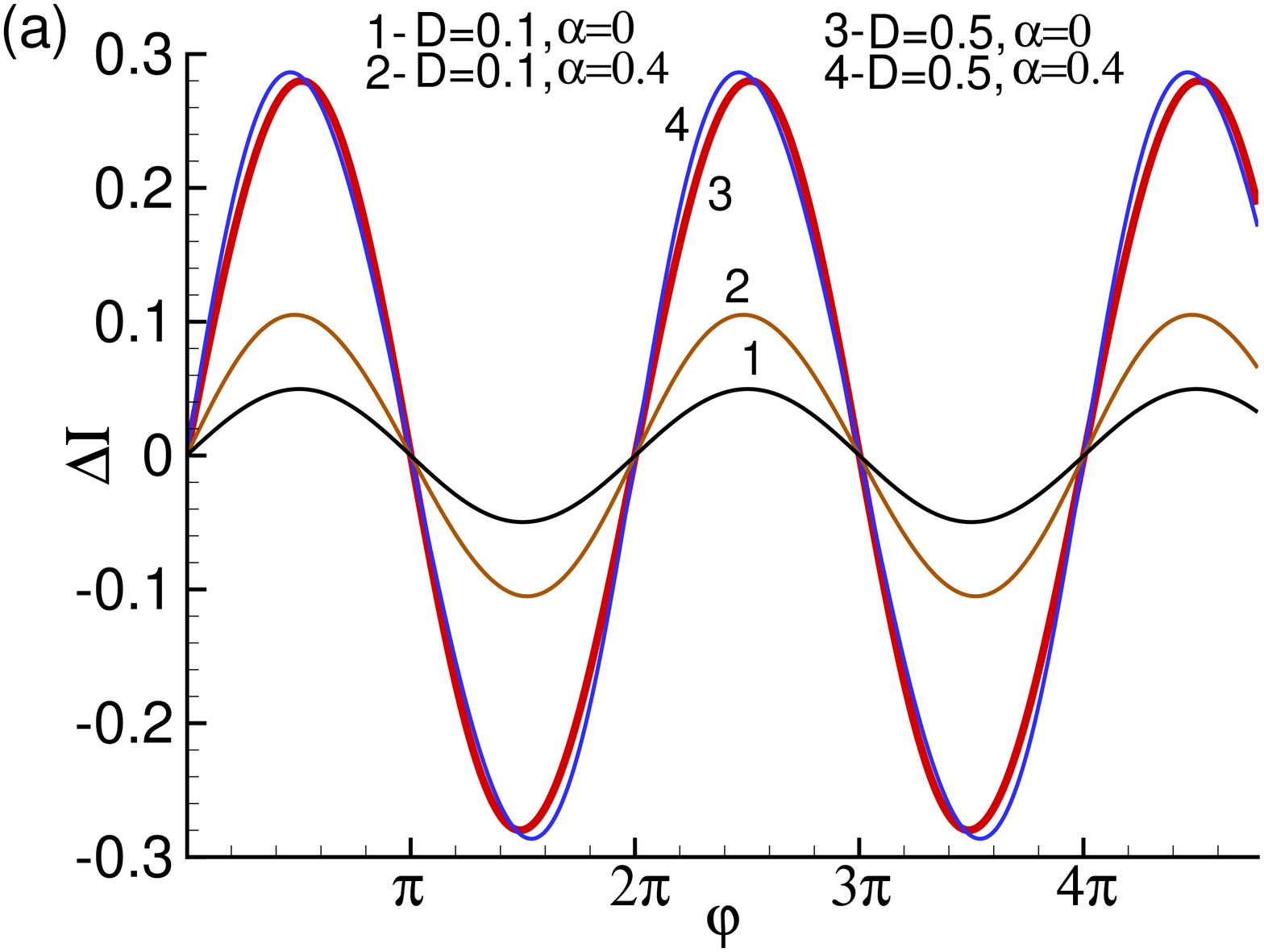}
\includegraphics[height=50mm]{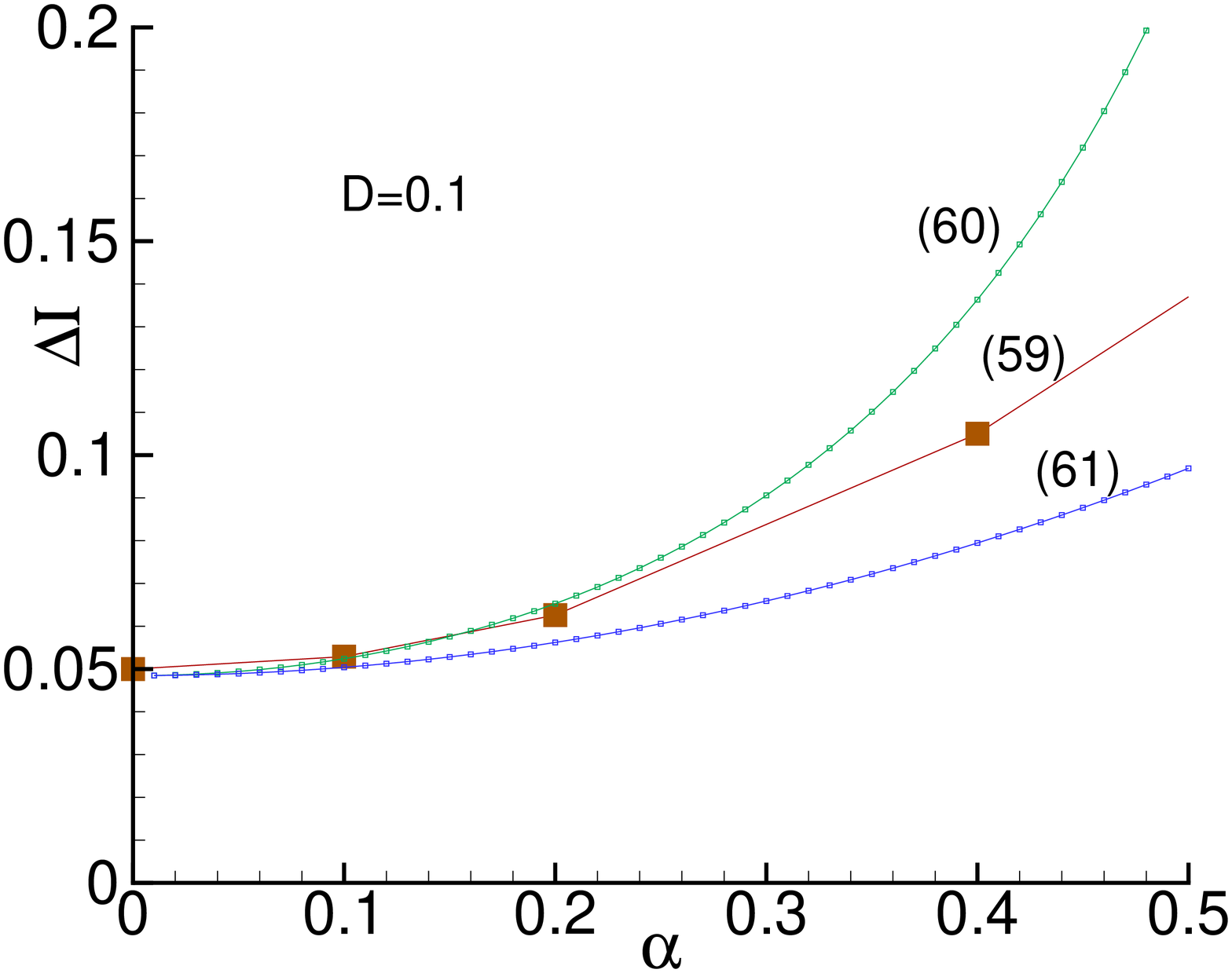}
\caption{(a) Effect of transparency and spin-orbital coupling on the
$\varphi$-dependence of the Shapiro step width according to the
formula (\ref{swidthso}); (b) Demonstration of $\alpha$-dependence of Shapiro
step width in different approximations according to the formulas
(\ref{swidthso}),   (\ref{swidthdll1}) and  (\ref{swidthdll2}).} \label{plots-width}
\end{figure}

Thus we find the step width depends on the magnitude of the
spin-orbit coupling. Indeed, Fig.\ \ref{plots-width}(a) demonstrates
this effect of transparency and spin-orbital coupling on the
$\varphi$-dependence of the Shapiro step width according to formula
(\ref{swidthso}).  We also note that for $D \ll 1$, the maxima and
minima of the DC current occur for $\varphi_0^{n_0 \alpha} \simeq
\pm \pi/2$ and Eq.\ (\ref{swidthso}) simplifies to yield
\begin{eqnarray}
\Delta I^{\alpha} (D\ll 1) &\simeq& \frac{ e \Delta D}{2\hbar}
J_{n_0}(\omega) \left( 2- \eta_{+} - \eta_{-}\right)
\label{swidthdll1}
\end{eqnarray}
For small $\tilde \alpha = \alpha/v_F$, it is easy to see by
expanding $\eta_{\pm}$ in power of $\tilde \alpha $, that
\begin{eqnarray}
\Delta I^{\alpha} (D\ll 1; \tilde \alpha \ll 1) &\simeq& \frac{ e
\Delta D}{\hbar} J_{n_0}(\omega) \left( 1+4\tilde\alpha^2 +
...\right) \label{swidthdll2} \nonumber\\
\end{eqnarray}
which demonstrates the dependence of step width on the SO coupling
$\alpha$.

Comparison of these three plots according to Eqs.\ (\ref{swidthso}),
(\ref{swidthdll1}) and  (\ref{swidthdll2}) is presented in  Fig.\
\ref{plots-width}(b). As we can see, the results of approximations
(\ref{swidthdll1}) and  (\ref{swidthdll2}) demonstrate more sharper
increasing of Shapiro step width with $\alpha$ in compare with
formula  (\ref{swidthso}). It's clear that the difference disappears
in the limit $D \rightarrow 0$. The obtained dependence of the SS
width on the spin-orbit coupling may be used for the experimental
estimation of its value.

\begin{figure}[h!]
 \centering
\includegraphics[height=60mm]{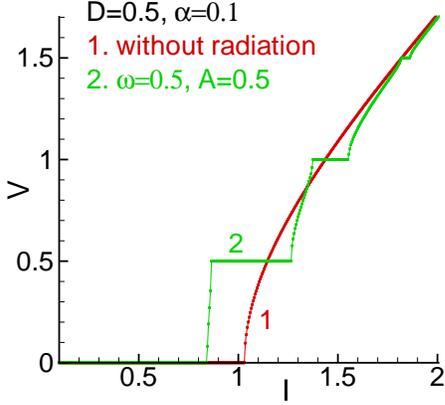}
\caption{I-V curve at $D=0.5$, $\alpha=0.1$ without radiation (curve
1) and under external radiation (curve 2)} \label{iv_curve}
\end{figure}

To investigate the effect of SOI on the amplitude dependence of
Shapiro step width, we have calculated the I-V curves for the
junction under external radiation using equation (\ref{acjos1}).
This result is presented in Fig.\ \ref{iv_curve}, where we show the
I-V curve of the junction at $D=0.5$, $\alpha=0.1$ under external
electromagnetic radiation with frequency $\omega=0.5$ and amplitude
$A=0.5$. In this figure we include for comparison the I-V
characteristics without radiation also. The I-V curve demonstrates
the main Shapiro step at $V=\omega=0.5$ and its harmonics.

\begin{figure}[h!]
 \centering
\includegraphics[height=50mm]{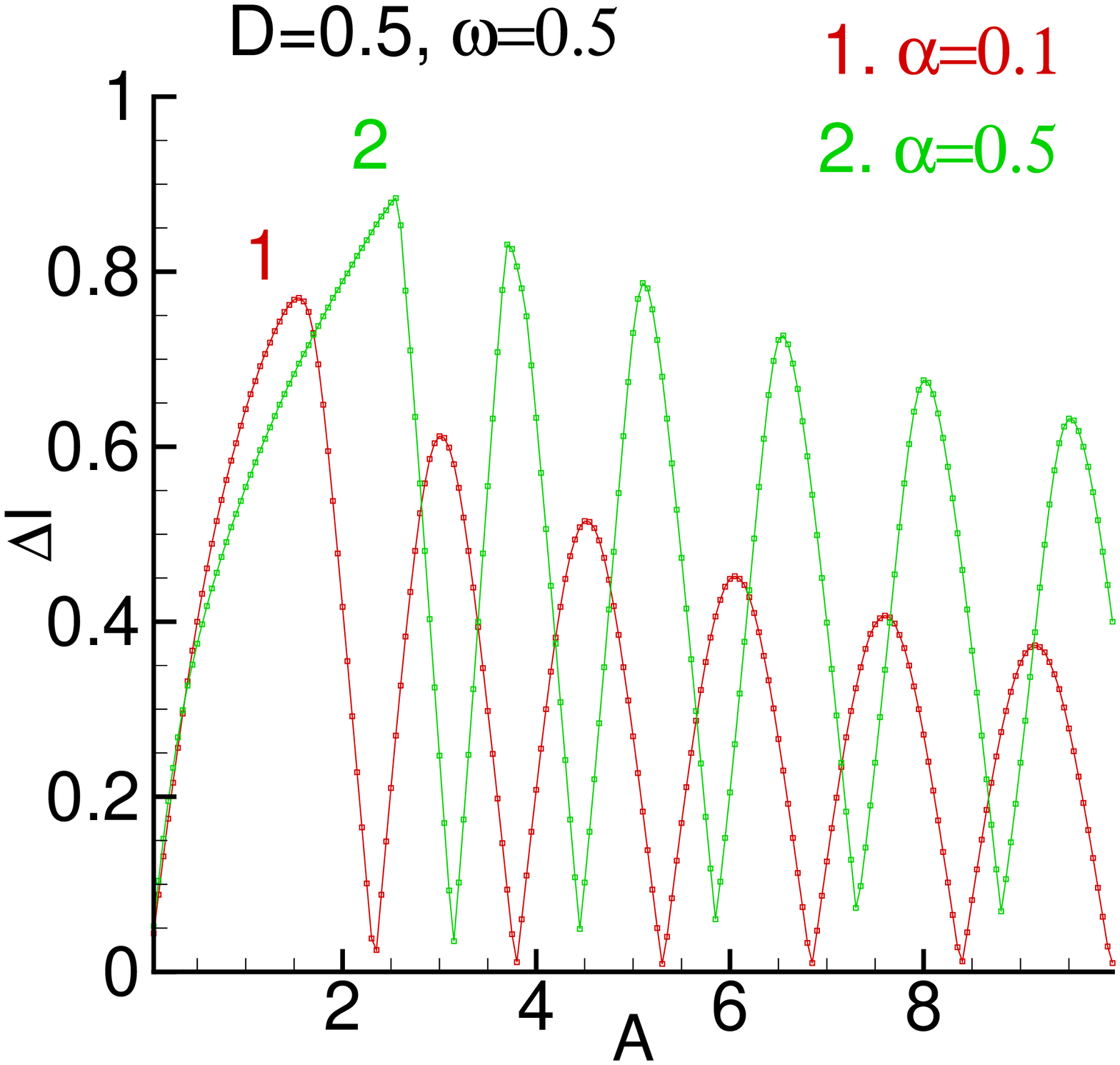}
\includegraphics[height=50mm]{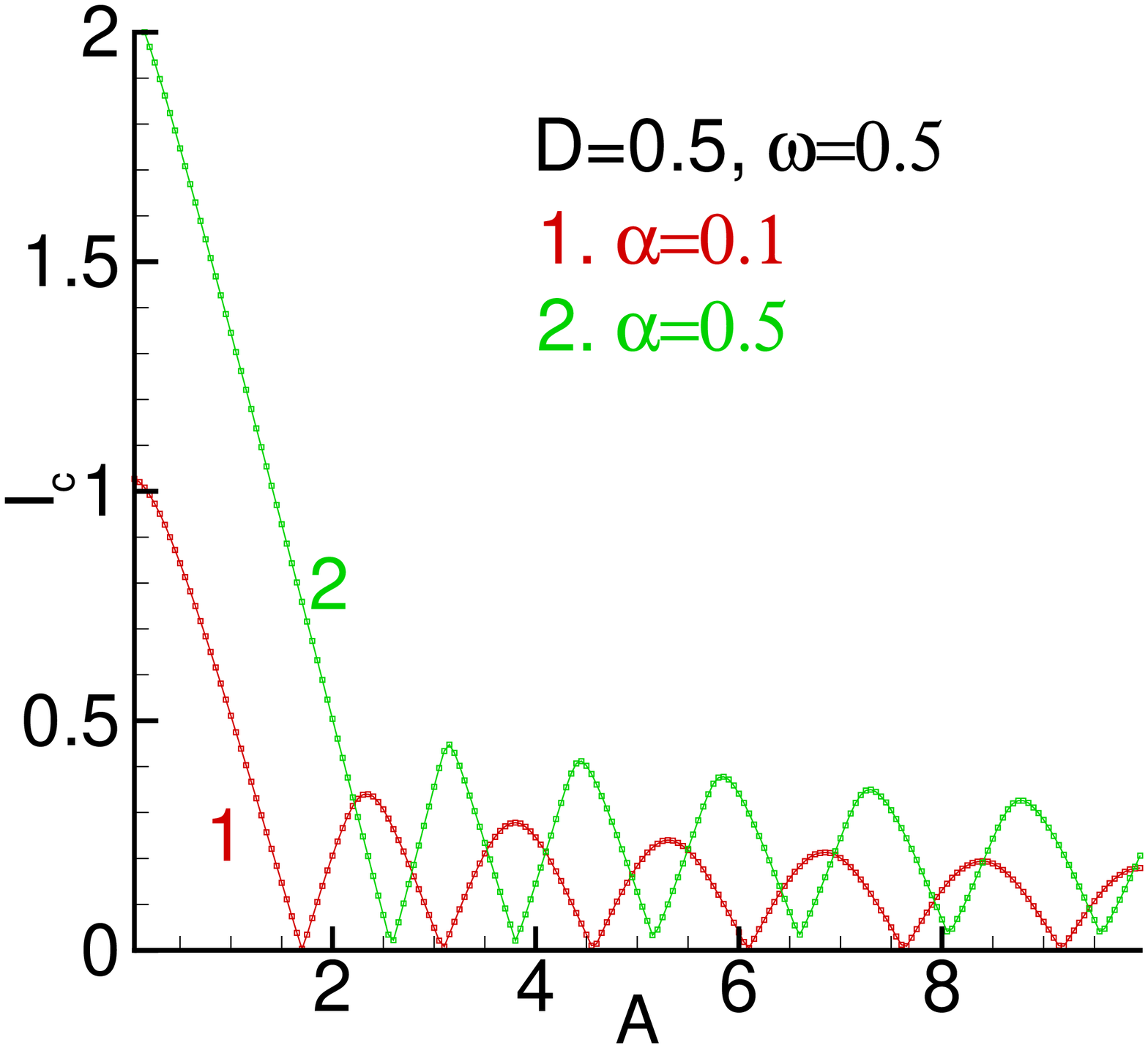}
\caption{Effect of spin orbital coupling on the amplitude dependence of:
(a) Shapiro step width; (b) Critical current.}
\label{amp_dep}
\end{figure}

Fig.\ \ref{amp_dep}(a) shows the amplitude dependence of Shapiro
step width in case $\alpha=0.5$ (line 1) and
 $\alpha=0.1$ (line 2) under external radiation with frequency
 $\omega=0.5$. Calculation is provided for value of
 transparency $D=0.5$. We see that the value of the SOI parameter has
 a noticeable effect on the Shapiro step width and its dependence
 on amplitude of the external radiation.  These results of I-V characteristics simulations   coincide qualitatively
 with the conclusion followed from Fig.\ \ref{plots-width}. We see that in  case
 with $\alpha=0.5$ the width of Shapiro step is larger than case
$\alpha=0.1$. The similar effect can be seen in amplitude dependence
of critical current $I_{c}$, which is shown in Fig.\
\ref{amp_dep}(b).

\begin{figure}[h!]
 \centering
\includegraphics[height=60mm]{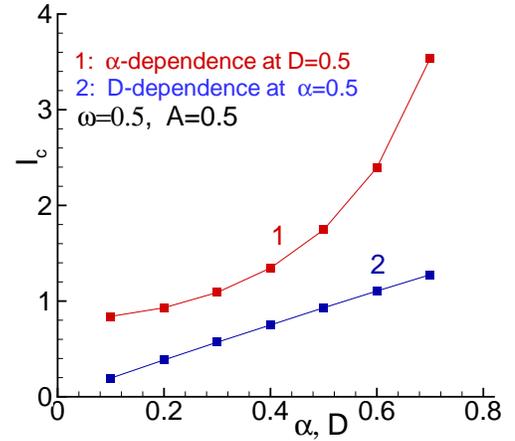}
\caption{The $\alpha$--dependence of $I_{c}$ for $D=0.5$ and
$D$--dependence of $I_{c}$ for $\alpha=0.2$ at $\omega=0.5$, $A=0.5$.}
\label{alpha_D-dep}
\end{figure}
The transparency coefficient $D$ also effects the critical current
value. To distinguish and clarify the effect of SOI we have
calculated the $\alpha$-- and $D$--dependence of $I_{c}$, which is
demonstrated in Figures~\ref{alpha_D-dep} (a) and (b). These results
might be used for the comparison with future experimental results.

\section{Conclusion}
\label{sec6}

In this paper we study the Josephson current between 1D
superconducting nanowires separated by an insulating barrier in the
presence of Rashba SOI and the magnetic fields ${\bf B}$ and $h$.
The presence of the SOI and Zeeman magnetic fields leads to four
distinct Fermi points in each bulk superconductor. Therefore, the
study of Josephson effect in these junctions requires construction
of an incident quasiparticle wave function which is in a linear
superposition state of plane waves around each Fermi points. In our
study, we have developed a theoretical method to study Josephson
effect in such systems; our work thus constitutes a generalization
of analysis of Ref.\ \onlinecite{ksy04} to systems with SOI and
Zeeman fields. We have provided analytical results for the Andreev
bound states in several asymptotic limits from our analysis,
demonstrated the presence of spin-Josephson current in these
junctions, and studied the dependence of Shapiro steps on SOI
interaction strength $\alpha$ in the presence of external radiation.
Moreover, we have demonstrated the existence of magneto-Josephson
effect in these systems. We note that although the existence of the
magneto-Josephson effect in a topological superconductor has been
predicted recently \cite{jpar13, kss12, pjpa13}, the question of
whether this effect is observable in  superconducting junctions with
quadratic electronic dispersion and the absence of SOI was not
addressed before. We show in the paper the magneto-Josephson effect
takes place even in the absence of SOI.

Experimental verification of our work would require experiments
conducted on Josephson junctions in 1D nanowires analogous to ones
studied in Ref.\ \onlinecite{rlf12}. We predict that the variation
of the angle $\phi$ of the in-plane magnetic field ${\bf B}$ would
lead to a spin-Josephson current as shown in Fig.\ \ref{47}.
Furthermore, AC Josephson effect measurement in these junction,
analogous to those done in Ref.\ \onlinecite{rlf12}, should reveal a
quadratic dependence of the Shapiro step-width as a function of
$\alpha$ for small $\alpha/v_F$ as shown in Fig.\ \ref{alpha_D-dep}.

Our work allows for several possible future direction. First, a
numerical solution of the condition ${\rm Det }\Lambda=0$ yielding
Andreev bound state energies in the regime where ${\bf B}, \alpha, h
\ne 0$ may lead to a better understanding of the interplay between
these parameters to shape the characteristics of the bound state
energies. Second, the formalism that we develop here may be extended
to regime of strong $\alpha$ and $B$ where the presence of Majorana
bound states shapes the characteristics of the Josephson current.
This requires a separate analysis since in this case the
quasiparticles would originates from two ( and not four Fermi
points) and is left as a topic for future study. Third, our
formalism may be applied to cases where the superconducting
pair-potential is unconventional (for example p-wave); indeed,
interplay of such unconventional pair-potentials and SO coupling may
lead to additional interesting characteristics in the Josepshon
current. We intend to explore these issues in future work.

In conclusion, we have studied Josephson effect in a unction between
two 1D nanowires in the presence of SOI and zeeman fields. We have
analyzed the Josephson current in these junctions and provided
analytical expressions of the Andreev bound states in several
limiting cases. We have also demonstrated the presence of
magneto-Josephson effect in these junctions and studied the Shapiro
step width in AC Josephson effect on the SOI strength. Our
theoretical predictions are shown to be verifiable by
straightforward experiments on these systems.

\section*{Acknowledgments}
The authors thank V. Osipov for discussion of this paper and
support. The reported study was funded partially by Azerbaijan-JINR
collaboration, the Science Development Foundation under the
President of the Republic Azerbaijan-Grant No
EIF-KETPL-2-2015-1(25)-56/01/1, the RFBR according to the research
projects 16--52--45011$\_$India, 15--51--61011$\_$Egypt, 15--29--01217 and DST-RFBR
grant.

\begin{widetext}
\appendix
\section{Energy dispersion for BdG superconductor}
\label{appa}

The expression ${\rm Det} |\mathcal{H} - E|=0$ for the energy
spectrum is written
\begin{equation*}
\begin{vmatrix}
E_{11}, & -Be^{-i \phi},& -\Delta , & 0\\
- Be^{i \phi} , & E_{22}, & 0 ,& \Delta\\
-\Delta^{\ast} , & 0 ,& E_{33}, & Be^{-i \phi}\\
0 , & \Delta^{\ast} , & Be^{i \phi} , & E_{44}
\end{vmatrix} =0
\end{equation*}
where $E_{11}=E+iabv_F k +iab \alpha k -h$,  $E_{22}=E+iabv_F k - iab \alpha k +h$,
$E_{33}=E - iabv_F k -iab \alpha k -h$, and  $E_{44}=E - iabv_F k + iab \alpha k + h$. Calculation of this
determinant  yields the energy spectrum of a ``bulk'' $1D$ superconductor

\begin{equation}
\left(E^2 -v_F^2k^2 + \alpha^2 k^2 -h^2-B^2 -|\Delta|^2 \right)^2 +4\left(Ekv_F + h \alpha k \right)^2 -
4 |\Delta|^2 \left(v_F^2 k^2 + B^2 + h^2 \right)=0
\label{Eo}
\end{equation}

This expression contains a linear in energy term, which is a result of an alignment of ${\bf h}$ and
the effective magnetic field of the SOI $\propto \alpha k$.

We consider different limiting cases below.
\begin{itemize}
\item{\bf The case of} $\alpha = B =h=0$.
\end{itemize}
The energy spectrum looks
\begin{equation}
E_{\pm}= \pm \sqrt{|\Delta|^2 -v_F^2 k^2}.
\label{E000}
\end{equation}
The energy levels of BdG quasi-particles lie in the gap, symmetrical to the Fermi level,
with momentum
\begin{equation}
k=\pm \frac{\sqrt{|\Delta|^2-E^2}}{v_F}.
\label{k000}
\end{equation}
\begin{itemize}
\item{\bf The case of} $B=0$, but $h \neq 0$ and $\alpha \neq 0$.
\end{itemize}
The energy spectrum (\ref{Eo}) in this limiting case is factorized

\begin{equation}
\left[(E+h)^2+(v_F-\alpha)^2k^2-|\Delta|^2\right]\left[(E-h)^2+(v_F+\alpha)^2k^2-|\Delta|^2\right]=0.
\label{E100}
\end{equation}
One gets for the quasi-particles' energy
\begin{equation}
E_{s,\pm}= sh \pm \sqrt{|\Delta|^2-(v_F + s \alpha )^2 k^2},
\label{E100}
\end{equation}
where $s=\pm$.
The momenta is expressed as
\begin{equation}
k_{\pm}^2=\frac{|\Delta|^2- (E \mp h)^2}{(v_F \pm \alpha)^2}.
\label{k100}
\end{equation}
SOI and/or magnetic field $h$ split both electron and  hole levels
due to Rashba 'momentum-shifting' and/or Zeeman effect. The 'Fermi
points' around $+k_F$ and $-k_F$ are split also due to these
effects.
\begin{itemize}
\item{\bf The limit of} $\alpha =0$, and  $B \neq 0$, $h \neq 0$.
\end{itemize}
Expression (\ref{Eo}) under these conditions reads
\begin{equation}
\left[\left(E+\sqrt{B^2+h^2}\right)^2+v_F^2k^2-|\Delta|^2\right]\left[\left(E-\sqrt{B^2+h^2}\right)^2+v_F^2k^2-|\Delta|^2\right]=0,
\end{equation}
yielding the following expression for the energy spectrum
\begin{equation}
E^2 = \left(\sqrt{|\Delta|^2 - v_F^2 k^2} \pm  \sqrt{ B^2 + h^2}\right)^2.
\label{E011}
\end{equation}
The momenta around the Fermi 'points' $+k_F$ and $-k_F$ split also
\begin{equation}
k_{\pm}^2=\frac{|\Delta|^2-\left(E \mp \sqrt{B^2+h^2}\right)^2}{v_F^2}.
\label{k011}
\end{equation}
The expressions for the energy and momentum in the limits of $\alpha =0$, ${\bf B}=0$ but $h \neq 0$ or
of $\alpha =0$, $h=0$ but ${\bf B} \neq 0$ are easily obtained from (\ref{E011}) and (\ref{k011}).
Note that a topological superconducting gapped phase is realized when $|\Delta|^2>B^2+h^2$ in consistent with
Ref.\cite{jpar13}.

\section{Computation of the Andreev bound states}
\label{appa2}

In this section, we chart out the expression for $\Lambda$.
The BdG wavefunction $\eta_a(x)$ can be written as a linear
superposition of its right and left moving components around
each Fermi momentum and with two different spins. Since we
look for bound state solutions, the general solution of
Eq.\ (\ref{Sch}) with (\ref{H2}) can be written as

\begin{widetext}
\begin{eqnarray}
\mathbf{\eta}_a(x)= \sum_{j= \pm} e^{{\rm sgn}(a) k_j x}\left[ A_a^j
\left(\begin{array}{ccc}
\eta_{a,\uparrow, +}(k_j) \\
\eta_{a, \downarrow, +}(k_j) \\
\eta_{a,\downarrow, -}^{\ast}(k_j) \\
\eta_{a, \uparrow, -}^{\ast}(k_j)
\end{array} \right)
e^{i k_{Fj} x} + B_a^j \left(\begin{array}{ccc}
\eta_{a,\uparrow, -}(k_j) \\
\eta_{a, \downarrow, -}(k_j) \\
\eta_{a,\downarrow, +}^{\ast}(k_j) \\
\eta_{a, \uparrow, +}^{\ast}(k_j)
\end{array} \right)
e^{-ik_{Fj} x} \right]
\label{wave}
\end{eqnarray}
\end{widetext}
where $k_a^{-1}$ denotes the localization length of the bound
states, and  ${\rm sgn}(a)=+(-)$ for $a=L(R)$. Henceforth, we shall
rename the coefficients as $A_a^+ \equiv A_a$, $A_a^- \equiv C_a$,
and $B_a^+ \equiv B_a$, $B_a^- \equiv D_a$ for clarity. Substituting
the wave functions (\ref{wave}) into the boundary conditions
(\ref{bc}) one gets eight linear homogeneous equations for $A_a$,
$B_a$, $C_a$, and $D_a$ with $a= \pm$ which can be represented in
terms of a $8 \times 8$ matrix $\Lambda$ and a column vector $\Phi =
( A_a, B_a, C_a, D_a)^T$ as $\Lambda \Phi =0$.
The energy of the Andreev bound states
can then be obtained from ${\rm Det} \Lambda =0$. The expression for
the matrix $\Lambda$, obtained from some straightforward algebra, is
given by
\begin{widetext}
\begin{eqnarray}
\Lambda &=& \left(\begin{array}{cc} D_1  & D_2 \\ D_3 & D_4
\end{array} \right) \nonumber\\
D_1 &=& \left(\begin{array}{cccc} \eta_{- \uparrow +}(k_+), &
\eta_{- \uparrow -}(k_+) ,& -\eta_{+ \uparrow -}(k_+), & -\eta_{+
\uparrow +}(k_+) \\ \eta_{- \downarrow +}(k_+), & \eta_{- \downarrow
-}(k_+)
,& -\eta_{+ \downarrow -}(k_+), & -\eta_{+ \downarrow +}(k_+) \\
\eta^{\ast}_{- \downarrow -}(k_+), & \eta^{\ast}_{- \downarrow
+}(k_+) ,& -\eta^{\ast}_{+
\downarrow +}(k_+), & -\eta^{\ast}_{+ \downarrow -}(k_+), \\
\eta^{\ast}_{- \uparrow -}(k_+), & \eta^{\ast}_{- \uparrow +}(k_+)
,& -\eta^{\ast}_{+ \uparrow +}(k_+), & -\eta^{\ast}_{+ \uparrow
-}(k_+)
\end{array} \right) \nonumber\\
D_2 &=& \left( \begin{array}{cccc} \eta_{- \uparrow +}(k_-), &
\eta_{- \uparrow -}(k_-), & -\eta_{+ \uparrow -}(k_-), & - \eta_{+
\uparrow +}(k_-) \\ \eta_{- \downarrow +}(k_-), & \eta_{- \downarrow
-}(k_-),
& -\eta_{+ \downarrow -}(k_-), & - \eta_{+ \downarrow +}(k_-) \\
\eta^{\ast}_{- \downarrow -}(k_-), & \eta^{\ast}_{- \downarrow
+}(k_-), & -\eta^{\ast}_{+ \downarrow +}(k_-), & - \eta^{\ast}_{+
\downarrow -}(k_-) \\ \eta^{\ast}_{- \uparrow -}(k_-), &
\eta^{\ast}_{- \uparrow +}(k_-), & -\eta^{\ast}_{+ \uparrow +}(k_-),
& - \eta^{\ast}_{+ \uparrow -}(k_-)
\end{array} \right) \nonumber\\
D_3 &=& \left( \begin{array}{cccc} C_{--}^+ \eta_{- \uparrow
+}(k_+), & -C_{++}^+\eta_{- \uparrow -}(k_+) , & C_{--}^+ \eta_{+
\uparrow -}(k_+), & -C_{++}^+ \eta_{+ \uparrow +}(k_+) \\
C_{--}^+\eta_{- \downarrow +}(k_+), & -C_{++}^+ \eta_{- \downarrow
-}(k_+) , & C_{--}^+ \eta_{+ \downarrow -}(k_+), & -C_{++}^+ \eta_{+
\downarrow +}(k_+) \\ C_{--}^+ \eta^{\ast}_{- \downarrow -}(k_+), &
-C_{++}^+ \eta^{\ast}_{- \downarrow +}(k_+) , & C_{--}^+
\eta^{\ast}_{+ \downarrow +}(k_+), & -C_{++}^+  \eta^{\ast}_{+
\downarrow -}(k_+) \\ C_{--}^+ \eta^{\ast}_{- \uparrow -}(k_+), &
-C_{++}^+ \eta^{\ast}_{- \uparrow +}(k_+) , & C_{--}^+
\eta^{\ast}_{+ \uparrow +}(k_+), & -C_{++}^+ \eta^{\ast}_{+ \uparrow
-}(k_+), \end{array} \right) \nonumber\\
D_4 &=& \left( \begin{array}{cccc} C_{--}^- \eta_{- \uparrow
+}(k_-), & -C_{++}^- \eta_{- \uparrow -}(k_-), & C_{--}^-\eta_{+
\uparrow -}(k_-), & -C_{++}^-\eta_{+ \uparrow +}(k_-)\\ C_{--}^-
\eta_{- \downarrow -}(k_-), & -C_{++}^- \eta_{- \downarrow -}(k_-),
& C_{--}^-\eta_{+ \downarrow -}(k_-), & -C_{++}^-\eta_{+ \downarrow
+}(k_-)\\ C_{--}^- \eta^{\ast}_{- \downarrow -}(k_-), & -C_{++}^-
\eta^{\ast}_{- \downarrow +}(k_-), & C_{--}^- \eta^{\ast}_{+
\downarrow +}(k_-), & -C_{++}^- \eta^{\ast}_{+ \downarrow -}(k_-)\\
C_{--}^- \eta^{\ast}_{- \uparrow -}(k_-), & -C_{++}^- \eta^{\ast}_{-
\uparrow +}(k_-), & C_{--}^- \eta^{\ast}_{+ \uparrow +}(k_-), &
-C_{++}^- \eta^{\ast}_{+ \uparrow -}(k_-) \end{array} \right)
\label{arrayeq1}
\end{eqnarray}
where $C_{\mu \nu}^{\pm} =(ik_F + \mu k_{\pm} +\nu k_FZ/2)$ and
$\nu, \mu$ takes values $\pm 1$.

We note that it is difficult to obtain analytical expression of
${\rm Det} \Lambda$ for general values of $B$, $\alpha$ and $h$.
However, the physical content of the several terms in this
determinant can be understood as follows. We define the minors of
the selected blocks of $\Lambda$ as ${\rm Det} D_1= {\tilde
F}^{\ast}_{\uparrow \downarrow}(k_+)$, ${\rm Det} D_2 = {\tilde
F}^{\ast}_{\uparrow \downarrow}(k_-)$, ${\rm Det} D_3 = {\tilde
F}_{\downarrow \uparrow}(k_+)$, ${\rm Det} D_4 = {\tilde
F}^{\ast}_{\downarrow \downarrow}(k_-)$. Furthermore we define the $
4 \times 4$ matrices
\begin{eqnarray}
D_5 &=& \left(\begin{array}{cccc} \eta_{- \uparrow +}(k_+), &
\eta_{- \uparrow -}(k_+) ,& -\eta_{+ \uparrow -}(k_+), & -\eta_{+
\uparrow +}(k_+) \\ \eta^{\ast}_{- \uparrow -}(k_+), &
\eta^{\ast}_{- \uparrow +}(k_+) ,& -\eta^{\ast}_{+ \uparrow +}(k_+),
& -\eta^{\ast}_{+ \uparrow -}(k_+) \\ C_{--}^+ \eta_{- \uparrow
+}(k_+), & -C_{++}^+\eta_{- \uparrow -}(k_+) , & C_{--}^+ \eta_{+
\uparrow -}(k_+), & -C_{++}^+ \eta_{+ \uparrow +}(k_+) \\ C_{--}^+
\eta^{\ast}_{- \uparrow -}(k_+), & -C_{++}^+ \eta^{\ast}_{- \uparrow
+}(k_+) , & C_{--}^+ \eta^{\ast}_{+ \uparrow +}(k_+), & -C_{++}^+
\eta^{\ast}_{+ \uparrow -}(k_+) \end{array} \right) \nonumber\\
D_6 &=&  \left(\begin{array}{cccc} \eta_{- \downarrow +}(k_+), &
\eta_{- \downarrow -}(k_+)
,& -\eta_{+ \downarrow -}(k_+), & -\eta_{+ \downarrow +}(k_+) \\
\eta^{\ast}_{- \downarrow -}(k_+), & \eta^{\ast}_{- \downarrow
+}(k_+) ,& -\eta^{\ast}_{+ \downarrow +}(k_+), & -\eta^{\ast}_{+
\downarrow -}(k_+) \\ C_{--}^+\eta_{- \downarrow +}(k_+), &
-C_{++}^+ \eta_{- \downarrow -}(k_+) , & C_{--}^+ \eta_{+ \downarrow
-}(k_+), & -C_{++}^+ \eta_{+ \downarrow +}(k_+) \\ C_{--}^+
\eta^{\ast}_{- \downarrow -}(k_+), & -C_{++}^+ \eta^{\ast}_{-
\downarrow +}(k_+) , & C_{--}^+ \eta^{\ast}_{+ \downarrow +}(k_+), &
-C_{++}^+  \eta^{\ast}_{+ \downarrow -}(k_+) \end{array} \right)
\label{arrayeq2}
\end{eqnarray}
\end{widetext}
The determinants of these matrices are denoted by ${\rm Det} D_5 =
F^{\ast}_{\uparrow \uparrow}(k_+)$ and ${\rm Det} D_6 =
F^{\ast}_{\downarrow \downarrow}(k_+)$. Similarly one can also
construct expressions for $F^{\ast}_{\uparrow \uparrow}(k_-)$ and
$F^{\ast}_{\downarrow \downarrow}(k_+)$. Note that all these blocks
are interpreted to correspond to a definitive physical process as
explained in the main text. All of these determinants enter the
expressions of the Andreev bound states as discussed in Sec.\
\ref{sec3} of the main text.

\section{Andreev bound states at {\bf B=0}}
\label{appb}

In this section we look into the expression of Andreev bound states
for $|{\bf B}|=0$. The boundary conditions (\ref{bc}) for the wave
function (\ref{wave}), written in the absence of the SOI induced
momentum splitting yield again eight equations for four coefficients
$A_{\pm}$ and $B_{\pm}$; these equations are BdG equations for a
s-wave superconductor with spin-dependent eigenfunctions
$\eta_{a,\sigma,b}$ and $\eta_{a,{\bar \sigma},b}^{\ast}$, where the
overline of an index (e.g., ${\bar \sigma}$) means an opposite
direction or sign. One chooses four equations corresponding to an
electron-hole pair with opposite spins. The determinant
corresponding to the matrix (defined as $D_1$ in Appendix\
\ref{appa2}) in the front of the coefficients $A_a$ and $B_a$  is
calculated to give
\begin{equation}
{\tilde F}_{\uparrow, \downarrow}^{\ast}=
\frac{1}{D^2}\eta_{+,\downarrow,+}
\eta_{-,\downarrow,-}\eta_{+,\downarrow,-}\eta_{-,\downarrow,+}
F^{\ast}_{\uparrow, \downarrow}, \label{Fdag}
\end{equation}
where
\begin{eqnarray}
F_{\uparrow, \downarrow}^{\ast}=\left[\frac{\eta^{\ast}_{-,
\uparrow,-}}{\eta_{-,\downarrow,+}} - \frac{\eta^{\ast}_{+,
\uparrow,-}}{\eta_{+,\downarrow,+}}\right]
\left[\frac{\eta^{\ast}_{+, \uparrow,+}}{\eta_{+,\downarrow,-}}
-\frac{\eta^{\ast}_{-, \uparrow,+}}{\eta_{-,\downarrow,-}}\right]-\\
\nonumber (1-D) \left[\frac{\eta^{\ast}_{+,
\uparrow,-}}{\eta_{+,\downarrow,+}} -\frac{\eta^{\ast}_{-,
\uparrow,+}}{\eta_{-,\downarrow,-}}\right]
\left[\frac{\eta^{\ast}_{-, \uparrow,-}}{\eta_{-,\downarrow,+}}
-\frac{\eta^{\ast}_{+, \uparrow,+}}{\eta_{+,\downarrow,-}}\right].
\label{energy0}
\end{eqnarray}
Equating this determinant to zero one gets a condition to find the
energy spectrum \cite{ksy04}. Note that the other four equations
yields the same expression with only spin being interchanged leading
to ${\tilde F}_{\downarrow, \uparrow}^{\ast}$.  It is easy to see
that the condition to determine the Andreev bound state energy in
this limit, where $\Lambda$ constitutes two $4 \times 4$ blocks, is
given by equating
\begin{eqnarray}
{\tilde F}_{\uparrow, \downarrow}^{\ast} \cdot {\tilde
F}_{\downarrow, \uparrow}^{\ast}  \label{ksy}
\end{eqnarray}
to zero. Eqs. (\ref{Sch1})..(\ref{Sch4}) allow us to calculate all
possible ratios $\eta^{\ast}_{a, \sigma, b}/\eta_{a,{\bar
\sigma},{\bar b}}$, $\eta^{\ast}_{a, \sigma, b}/\eta_{a,\sigma,
{\bar b}}$, and $\eta_{a, \sigma,b}/\eta_{a,{\bar \sigma}, b}$,
$\eta^{\ast}_{a, \sigma,b}/\eta^{\ast}_{a,{\bar \sigma}, b}$.
Furthermore, we note that only the ratio $\eta^{\ast}_{a, \sigma,
b}/\eta_{a,{\bar \sigma},{\bar b}}$ is non-zero for ${\bf B}=0$. We
shall return to this case below.

Next, we note from Eqs. (\ref{Sch1})..(\ref{Sch4}) that the
dependencies of these equations on $\phi$ and $\varphi$ are
completely removed by transforming the wave function as
\begin{eqnarray}
\eta^{\ast}_a(x) &\to& \left(e^{-i(\varphi-
\phi)/2}\eta_{a,\uparrow, \bar{b}}^{\ast}(x), ~e^{-i(\varphi
+\phi/2)} \eta_{a,
\downarrow,\bar{b}}^{\ast}(x), \right. \nonumber\\
&& \left.~ e^{i(\varphi+ \phi)/2}\eta_{a, \downarrow, b} (x),
~e^{i(\varphi- \phi)/2} \eta_{a, \uparrow, b}(x)\right).
\end{eqnarray}
In the transformed basis one has
\begin{eqnarray}
\frac{\eta^{\ast}_{a, \uparrow, b}}{\eta_{a,\uparrow,{\bar b}}} \to
e^{-i(\varphi - \phi)} \frac{\eta^{\ast}_{a, \uparrow,
b}}{\eta_{a,\uparrow, {\bar b}}}, \quad \frac{\eta^{\ast}_{a,
\downarrow, b}}{\eta_{a,\downarrow, {\bar b}}} \to e^{-i(\varphi +
\phi)} \frac{\eta^{\ast}_{a, \downarrow,
b}}{\eta_{a,\downarrow, {\bar b}}} \label{dagup-up} \\
\frac{\eta^{\ast}_{a, \uparrow, b}}{\eta_{a,\downarrow, {\bar b}}}
\to e^{-i\varphi} \frac{\eta^{\ast}_{a, \uparrow,
b}}{\eta_{a,\downarrow, {\bar b}}}, \quad   \frac{\eta^{\ast}_{a,
\downarrow, b}}{\eta_{a,\uparrow, {\bar b}}} \to e^{-i\varphi}
\frac{\eta^{\ast}_{a, \downarrow,
b}}{\eta_{a,\uparrow, {\bar b}}} \label{dagup-down} \\
\frac{\eta^{\ast}_{a, \uparrow, b}}{\eta_{a,\downarrow, b}^{\ast}}
\to e^{i\phi} \frac{\eta^{\ast}_{a, \uparrow,
b}}{\eta_{a,\downarrow, b}^{\ast}}, \quad  \frac{\eta_{a, \uparrow,
b}}{\eta_{a,\downarrow, b}} \to e^{-i\phi} \frac{\eta_{a, \uparrow,
b}}{\eta_{a,\downarrow, b}}. \label{up-up}
\end{eqnarray}
The different ratios that appear in the left-side of Eqs.\
\ref{dagup-up}..\ref{up-up} can be understood as follows. The ratio
$\eta^{\ast}_{a, \sigma, b}/\eta_{a,{\bar \sigma},{\bar b}}$
corresponds to the amplitude of conventional Andreev reflection
channel which constitutes reflection of an electron-like
quasiparticle to a hole-like quasiparticle with opposite spin on a
N-S interface. In contrast, the ratio $\eta^{\ast}_{a, \sigma,
b}/\eta_{a,\sigma, {\bar b}}$ which is finite only in the presence
of SOI and/or magnetic field, represents amplitude of Andreev
reflection channel where the electron-like quasiparticle incident on
the interface is reflected to a hole-like quasiparticle state with
the same spin orientation. Finally, the ratio $\eta_{a,
\sigma,b}/\eta_{a,{\bar \sigma},} b$ represents a usual reflection
channel of an electron-like quasiparticle on the boundary without
creation of a Cooper pair in a superconducting part of the junction.
Since these ratios enter the expressions of $F_{\sigma \sigma'}$,
these also represents Andreev and normal reflection processes
involving electron-like and hole-like quasiparticles in the opposite
($\sigma'= \bar \sigma$) and same ($\sigma'=\sigma$) spin sector. We
note that the ratio of wavefunctions in Eq.\ (\ref{dagup-up}) depend
on both $\phi$ and $\varphi$ while those in Eqs. (\ref{dagup-down})
and (\ref{up-up}) depend on either $\varphi$ or $\phi$. This
suggests that the ratios (\ref{dagup-up}) and (\ref{dagup-down}) are
responsible for the dependence of observable parameters on the order
parameter phase difference $\varphi$, whereas the ratios
(\ref{dagup-up}) and (\ref{up-up}) are responsible for the
dependence on the  magnetic field orientation angle $\phi$.

The ratios $\eta^{\ast}_{a, \uparrow,{\bar
b}}/\eta_{a,\downarrow,b}$ and $\eta^{\ast}_{a, \downarrow,{\bar
b}}/\eta_{a,\uparrow,b}$ are determined from Eqs.
(\ref{Sch1})-(\ref{Sch4}) as
\begin{eqnarray}
\frac{\eta^{\ast}_{a, \sigma,\bar{b}}}{\eta_{a,\bar{\sigma},b}} &=&
\pm \frac{1}{\Delta_a} \bigg\{ \frac{M_{\pm}(k)}{2(E \pm iab \alpha
k)} \nonumber\\
&& -(E+i a b v_F k \mp i ab \alpha k \pm h)\bigg\},
\label{wave-up-down} \\
M_{\pm}(k)&=& (E \pm h)^2+(v_F k \mp \alpha k)^2 + B^2 - |\Delta|^2,
\label{M}
\end{eqnarray}
where the upper (lower) sign $+$ ($-$) corresponds to spin $\sigma =
\uparrow$ ($\sigma = \downarrow$). Using Eq.\ (\ref{wave-up-down}),
one obtains, after a few lines of algebra, the expressions for
$F_{\uparrow, \downarrow}^{\ast}$ and $F_{\downarrow,
\uparrow}^{\ast}$ for  general ${\bf B}$, $h$ and $\alpha$ as
\begin{eqnarray}
&& F_{\sigma {\bar \sigma}}^{\ast}(k) = \big\{ \big [\alpha k
M_{\pm}(k) \pm 2(v_F k \mp \alpha k)(E^2 +\alpha^2 k^2)\big]^2 \nonumber\\
&& -4D|\Delta|^2 (E^2+ \alpha^2 k^2)^2 \sin^2
\frac{\varphi}{2}\big\} (|\Delta|^2(E^2+\alpha^2 k^2)^2)^{-1},
\label{F-up-down} \nonumber\\
\end{eqnarray}

Equations (\ref{Sch1})-(\ref{Sch4}) are strongly simplified in this
link providing only a link between $\eta_{a,\sigma, b}$ and
$\eta^{\ast}_{a, {\bar \sigma}, {\bar b}}$
\begin{widetext}
\begin{eqnarray}
&&\frac{\eta^{\ast}_{a, \downarrow,{\bar
b}}}{\eta_{a,\uparrow,b}}=\frac{E+iabv_Fk + iab \alpha k
-h}{\Delta}= \frac{\Delta^{\ast}}{E -iab v_F k-iab \alpha k -h};
\label{Sch1B0}\\
&&\frac{\eta^{\ast}_{a, \uparrow,{\bar b}}}{\eta_{a,\downarrow,b}}=-
\frac{E+iab v_Fk -iab \alpha k +h}{\Delta}= -\frac{\Delta^{\ast}}{E
- iabv_Fk +iab \alpha k +h}. \label{Sch2B0}
\end{eqnarray}
Then, one gets for $F_{\uparrow, \downarrow}^{\ast}$ according to
Eq.\ (\ref{energy0})
\begin{equation}
F_{\uparrow, \downarrow}^{\ast}(k)=
\frac{4}{|\Delta|^2}\left\{(v_F+\alpha)^2k^2 -D [(E-h)^2 + (v_F
+\alpha)^2k^2] \sin^2\frac{\varphi}{2}\right\}. \label{ApF}
\end{equation}
\end{widetext}
The expression for $F_{\downarrow, \uparrow}^{\ast}(k)$ differs from
that for $F_{\uparrow, \downarrow}^{\ast}(k)$ by replacing $\alpha
\to -\alpha$ and $h \to -h$ in Eq.\ (\ref{ApF}). In the absence of
the magnetic fields a contribution to the bound energy due to SOI
comes from the {\it 'conventional'}  Andreev reflection connecting
electron-like and hole-like quasiparticles with opposite spins.
These can be expressed as
\begin{equation}
{\tilde F}_{\uparrow, \downarrow}^{\ast}(k_+) {\tilde
F}_{\downarrow, \uparrow}^{\ast}(k_-)  - {\tilde F}_{\uparrow,
\downarrow}^{\ast}(k_-) {\tilde F}_{\downarrow,
\uparrow}^{\ast}(k_+), \label{cont1}
\end{equation}
where ${\tilde F}_{\sigma {\bar \sigma}}$ can be obtained using
Eqs.\ \ref{F-up-down} and \ref{energy0}. In contrast, the main
tunneling channel in the presence of the magnetic field constitutes
an electron-like quasiparticle with a given spin polarization being
Andreev reflected to a hole-like quasiparticle state with the same
spin. The contribution to the bound state energy from this channel
is
\begin{eqnarray}
{\tilde F}_{\uparrow, \uparrow}^{\ast}(k_+) {\tilde F}_{\downarrow,
\downarrow}^{\ast}(k_-)  - {\tilde F}_{\uparrow,
\uparrow}^{\ast}(k_-) {\tilde F}_{\downarrow,
\downarrow}^{\ast}(k_+) \label{cont2}
\end{eqnarray}
where ${\tilde F}_{\sigma \sigma}^{\ast}(k)$ is given by
\begin{eqnarray}
{\tilde F}_{\sigma \sigma}^{\ast}(k) &=&
\frac{1}{D^2}\eta_{+,\sigma,+}(k) \eta_{-,\sigma,-}(k) \nonumber\\
&& \times  \eta_{+,\sigma,-}(k) \eta_{-,\sigma,+}(k)
F^{\ast}_{\sigma \sigma}(k). \label{feqss}
\end{eqnarray}

We note that $F_{\uparrow \uparrow}^{\ast}(k)$ (or $F_{\downarrow
\downarrow}^{\ast}(k)$) in Eq.\ (\ref{feqss}) is determined by Eq.\
(\ref{energy0}) after replacing the ratio  $\eta^{\ast}_{a, \sigma,
b}/\eta_{a,{\bar \sigma}, {\bar b}}$ in  $F_{\sigma, {\bar
\sigma}}^{\ast}$ by $\eta^{\ast}_{a, \sigma, b}/\eta_{a,\sigma,
{\bar b}}$. The expressions for $\eta^{\ast}_{a, \sigma,
b}/\eta_{a,\sigma, {\bar b}}$ can be obtained from Eqs.
(\ref{Sch1})-(\ref{Sch4})
\begin{eqnarray}
\frac{\eta^{\ast}_{a, \sigma, b}}{\eta_{a,\sigma, {\bar b}}} &=& \pm
\left\{ B e^{\pm i \phi_a}- (E+i a b v_F k \mp i
ab \alpha k \pm h) \right. \nonumber\\
&& \left. \times \eta_{a, \sigma, b}/\eta_{a,{\bar
\sigma}, b} \right\}/\Delta_a , \label{wave-r1} \\
\frac{\eta_{a, \sigma, b}}{\eta_{a, {\bar \sigma},b}} &=&
M_{\pm}(k)/(2 B e^{\pm i \phi_a} (E \pm i ab \alpha k)),
\label{wave-r2}
\end{eqnarray}
where the upper(lower) signs correspond to $\sigma =
\uparrow(\downarrow)$. These ratios can be used to obtain $F_{\sigma
\sigma}^{\ast}(k)$ as
\begin{eqnarray}
F_{\sigma \sigma}^{\ast}(k) &=& \frac{16 B^2}{|\Delta|^2
M^2_{\pm}(k)}\left\{\left(Ev_F k+\alpha k h\right)^2- D |\Delta|^2
\right.\nonumber\\
&& \left.\times \left(E^2+\alpha^2 k^2\right) \sin^2 \frac{\varphi
\mp \phi}{2}\right\}. \label{energyM}
\end{eqnarray}

Finally, the contribution to the bound energy from the channel given
by (\ref{up-up}) can be  expressed as
\begin{equation}
{\tilde F}_{\uparrow, \downarrow}(k_+) {\tilde F}_{\downarrow,
\uparrow}^{\ast~\ast}(k_-) - {\tilde F}_{\uparrow, \downarrow}(k_-)
{\tilde F}_{\downarrow, \uparrow}^{\ast~\ast}(k_+), \label{cont3}
\end{equation}
where
\begin{eqnarray}
{\tilde F}_{\sigma, {\bar \sigma}}(k) &=& \frac{1}{D^2}\eta_{+,{\bar
\sigma},+} \eta_{-,{\bar \sigma},-} \eta_{+, {\bar
\sigma},-}\eta_{-,{\bar \sigma},+} F^{\ast}_{\sigma, {\bar \sigma}}.
\end{eqnarray}
A procedure, similar to the one outlined above yields
\begin{eqnarray}
F_{\uparrow, \downarrow} &=& \left[\frac{\eta_{-,
\uparrow,+}}{\eta_{-,\downarrow,+}} -\frac{\eta_{+,
\uparrow,+}}{\eta_{+,\downarrow,+}}\right] \left[\frac{\eta_{+,
\uparrow,-}}{\eta_{+,\downarrow,-}}  - \frac{\eta_{-,
\uparrow,-}}{\eta_{-,\downarrow,-}}\right] \nonumber\\
&& -(1-D) \left[\frac{\eta_{+, \uparrow,+}}{\eta_{+,\downarrow,+}}
-\frac{\eta_{-, \uparrow,-}}{\eta_{-,\downarrow,-}}\right]
\left[\frac{\eta_{-, \uparrow,+}}{\eta_{-,\downarrow,+}}
-\frac{\eta_{+, \uparrow,-}}{\eta_{+,\downarrow,-}}\right]
\nonumber\\
&=& \frac{16 B^2}{M^2_-(k)}\left[\alpha^2 k^2-D(E^2 +\alpha^2 k^2)
\sin^2\frac{\phi}{2}\right]. \label{Fmag2}
\end{eqnarray}
The expression for  $F^{\ast ~\ast}_{\uparrow \downarrow}(k)$
differs from $F_{\uparrow \downarrow}(k)$ by replacing $M_-(k) \to
M_+(k)$ in (\ref{Fmag2}).

By equating to zero the sum of the expressions (\ref{ksy}),
(\ref{cont1}), (\ref{cont2}), and (\ref{cont3}) yields the Andreev
bound state energy in the presence of SOI and magnetic fields. In
what follows, we shall discuss two limiting case where a simple
analytical expressions for these bound states can be obtained.

The tunneling energy in this case receives its contribution from the
expression
\begin{widetext}
\begin{equation}
\left[v_F^2 k^2-D \sin^2 \frac{\varphi}{2}\right]^2 + F_{\uparrow,
\downarrow}^{\ast}(k_+) F_{\downarrow, \uparrow}^{\ast}(k_-) -
F_{\uparrow, \downarrow}^{\ast}(k_-) F_{\downarrow,
\uparrow}^{\ast}(k_+)=0 \label{ApFeq}
\end{equation}
with energy spectrum obtained from Eq.\ (\ref{Sch1B0})
\begin{equation}
(E-h)^2+(v_F + \alpha)^2k_+^2-|\Delta|^2=0 \quad {\rm and} \quad k_+^2=\frac{|\Delta|^2-(E-h)^2}{(v_F+\alpha)^2}
\label{Apk+}
\end{equation}
and from Eq.\ (\ref{Sch2B0})
\begin{equation}
(E+h)^2+(v_F - \alpha)^2k_-^2-|\Delta|^2=0 \quad {\rm and} \quad k_-^2=\frac{|\Delta|^2-(E+h)^2}{(v_F-\alpha)^2}.
\label{Apk-}
\end{equation}
\end{widetext}
This expression has been used to analyze Eq.\ (\ref{energyB=0}) of the
main text.
\end{widetext}

%


\begin{thebibliography}{99}

\bibitem{kitaev03} A. Yu. Kitaev, Annals Phys. {\bf 303}, 2 (2003).
\bibitem{nssf08} C. Nayak, S. Simon, A. Stern, M. Freedman, and S. Das Sarma, Rev. Mod. Phys. {\bf 80}, 1083 (2008).
\bibitem{alicea12} J. Alicea, Rep. Prog. Phys. {\bf 75}, 076501 (2012).
\bibitem{mzfp12} V. Mourik, K. Zuo, S. M. Frolov, S. R. Plissard, E. P. A. M. Bakkers, and
L. P. Kouwenhoven, Science {\bf 336}, 1003 (2012).
\bibitem{sen1}K. Sengupta, I. Zutic, H.J. Kwon, V.M. Yakovenko, S Das sarma
Physical Review B {\bf 63 (14)}, 144531 (2001).
\bibitem{ksy04} H. -J. Kwon, K. Sengupta, and V. M. Yakovenko, Eur. Phys. J. B {\bf 37}, 349 (2004).
\bibitem{drmo12} A. Das, Y. Ronen, Y. Most, Y. Oreg, M. Heiblum, and H. Shtrikman,  Nature Phys. {\bf 8},
887 (2012).
\bibitem{rlf12} L. P. Rokhinson, X. Liu, and J. K. Furdyna, Nature Phys. {\bf 8}, 795 (2012).
\bibitem{dyhl12} M. T. Deng, C. L. Yu, G. Y. Huang, M. Larsson, P. Caroff, and H. Q. Xu,
Nano Lett. {\bf 12}, 6414 (2012).
\bibitem{cfgd13} H. O. H. Churchill, V. Fatemi, K. Grove-Rasmussen, M. T. Deng, P. Caroff,and
C. M. Markus, Phys. Rev. B {\bf 87}, 241401 (R) (2013).
\bibitem{fhmj13} A. D. K. Finck, D. J. Van Harlingen, P. K. Mohseni, K. Jung, and X. Li,
Phys. Rev. Lett. {\bf 110}, 126406 (2013).
\bibitem{lsd10} R. M. Lutchyn, J. D. Sau, and S. Das Sarma, Phys. Rev. Lett. {\bf 105}, 077001 (2010).
\bibitem{oro10} Y. Oreg, G. Refael, and F. von Oppen, Phys. Rev. Lett. {\bf 105}, 177002 (2010).
\bibitem{kitaev01} A. Kitaev,  Phys. Usp. {\bf 44}, 131 (2001).
\bibitem{nab08} J. Nilsson, A. R. Akhmerov, and C. W. J. Beenakker, Phys. Rev. Lett. {\bf 101}, 120403 (2008).
\bibitem{fk08} L. Fu and C. L. Kane, Phys. Rev. Lett. {\bf 100}, 096407 (2008).
\bibitem{fk09} L. Fu and C. L. Kane, Phys. Rev. B {\bf 79}, 161408 (2009).
\bibitem{bhm11} D. M. Badiane, M. Houzet, and J. S. Meyer, Phys. Rev. Lett. {\bf 107}, 177002 (2011).
\bibitem{jpar11} L. Jiang, D. Pekker, J. Alicea, G. Refael, Y. Oreg, and F. von Oppen,
Phys. Rev. Lett. {\bf 107}, 236401 (2011).
\bibitem{spa12} P. San-Jose, E. Prada, and R. Aguado, Phys. Rev. Lett. {\bf 108}, 257001 (2012).
\bibitem{pn12} D. I. Pikulin and Y. V. Nazarov, Phys. Rev. B {\bf 86}, 140504 (2012).
\bibitem{ojanen13} T. Ojanen, Phys. Rev. B {\bf 87}, 100506(R) (2013).
\bibitem{scpa13} P. San-Jose, J. Cayao, E. Prada, and R. Aguado, New Journal of Physics {\bf 15}, 075019 (2013).
\bibitem{lmay14} S. -P. Lee, K. Michaeli, J. Alicea, and A. Yacoby, Phys. Rev. Lett. {\bf 113}, 197001 (2014).
\bibitem{clgt15} G. Campagnano, P. Lucignano, D. Giuliano, and A. Tagliacozzo, J. Phys.: Condens. Matter
{\bf 27}, 205301 (2015).
\bibitem{jpar13} L. Jiang, D. Pekker, J. Alicea, G. Refael, Y. Oreg, A. Brataas, and F. von Oppen,
Phys. Rev. B {\bf 87}, 075438 (2013); S. Jacobsen, I. Kulagina, and
J. Linder, Sci. Rep. 6, 23926 (2016).
\bibitem{kss12} P. Kotetes, G. Sch\"on,  and A. Shnirman, J. Korean Phys. Soc. {\bf 62}, 1558 (2013).
\bibitem{bphs13} C. W. J. Beenakker, D. I. Pikulin, T. Hyart, S. Schomerus, and J. P. Dahlhaus,
Phys. Rev. Lett. {\bf 110}, 017003 (2013).
\bibitem{pjpa13} F. Pientika, L. Jiang, D. Pekker, J. Alicea, G. Refael, Y. Oreg, and F. von Oppen,
New J. Physics {\bf 15}, 115001 (2013).
\bibitem{zk14} F. Zhang and C. L. Kane, Phys. Rev. Lett. {\bf 113}, 036401 (2014).
\bibitem{abold} I.O. Kulik, Zh. Eksp. Teor. Fiz. {\bf 57}, 1745 (1969) [Sov. Phys.
JETP {\bf 30}, 944 (1970)]; C. Ishii, Progr. Theor. Phys. {\bf 44},
1525 (1970); J. Bardeen, J.L. Johnson, Phys. Rev. B 5, {\bf 72}
(1972); T. L¨ofwander, V.S. Shumeiko, G. Wendin, Supercond. Sci.
Technol. {\bf 14}, R53 (2001).
\bibitem{zag1} See for example A.M. Zagoskin, {\it Quantum Thheory
of Many-Body Systems: Techniques and Applications}, Springer-Verlag,
New York (1998).

\end{thebibliography}
\end{document}